\ifdefined\XeTeXversion\else\pdfoutput=1\fi  
\ifdefined\XeTeXversion\else\pdfminorversion=7\fi
\documentclass[preprint,10pt,number,sort&compress,a4paper]{elsarticle}
\usepackage[T1]{fontenc}
\usepackage[utf8]{inputenc}
\usepackage{amsmath,amssymb,bm}
\usepackage{graphicx}
\usepackage{booktabs}
\usepackage{siunitx}
\usepackage[colorlinks=true,linkcolor=blue,citecolor=blue,urlcolor=blue]{hyperref}
\usepackage{xurl}
\usepackage{float}
\usepackage[section]{placeins}
\graphicspath{{figs/}}
\newcommand{\cb}{\bar{c}}
\newcommand{\Pf}{P_f}
\newcommand{\dd}{\mathrm{d}}
\journal{Acta Geotechnica}

\begin{document}
\begin{frontmatter}

\title{Label-free physics-informed strength reduction and a neural operator for the reliability of spatially variable slopes}

\author{Apisit Robjanghvad}
\author{Sompote Youwai\corref{cor1}}
\cortext[cor1]{Corresponding author. Email: sompote.you@kmutt.ac.th}
\address{AI Research Group, Department of Civil Engineering, Faculty of Engineering, King Mongkut's University of Technology Thonburi, Bangkok, Thailand}

\begin{abstract}
This paper proposes a label-free physics-informed neural network (PINN) for non-associated Mohr--Coulomb plasticity with incremental strength reduction for the reliability of spatially variable $c$--$\varphi$ slopes. The PINN is trained with a virtual-work loss on the finite-element mesh and solves many random fields at once on a graphics processing unit. With a solver budget fixed on the homogeneous slopes, it reproduces the Abaqus factor of safety within 2 to 3\,\% root-mean-square on 52 random fields and locates the failure mechanism in the same weak zones, with the plastic strain in the band under-resolved. A neural operator trained only on the PINN solutions predicts the factor of safety, displacement path and failure mode of a million fields in seconds. No finite-element solution enters the pipeline: Monte Carlo fields, importance samples and operator training data come from the PINN, and Abaqus serves only to verify the result. Monte Carlo simulation, the first-order reliability method, importance sampling and the operator give consistent failure probabilities down to $10^{-3}$ for the slope studied, but a 3\,\% scatter in the factor of safety inflates a probability of $10^{-3}$ by up to a factor of two, and the absolute values refer to one mesh. A heterogeneous Bishop analysis gives failure probabilities 2.6 and 8.3 times lower although its critical circles follow the PINN mechanism, because a small difference in the factor of safety is amplified in the tail. The operator also yields the failure mechanism and displacement response conditional on failure.
\end{abstract}

\end{frontmatter}

\section{Introduction}\label{sec:intro}

Natural soils exhibit spatial variability in shear strength, and slope stability is governed not only by the mean strength but also by the spatial arrangement of weak zones. Probabilistic slope stability analysis therefore represents the strength parameters as random fields. The random finite-element method (RFEM) combines such fields with elastoplastic finite-element analysis and the strength reduction method (SRM) \citep{griffiths1999,dawson1999} in a Monte Carlo framework \citep{griffiths2004}. In RFEM, the failure mechanism is not prescribed but emerges along the weakest path in each realisation, while slope failure is treated as a system event \citep{huang2010}. Analyses that neglect spatial variability can yield unconservative estimates of failure probability \citep{griffiths2009}, while a slope classified as safe in a deterministic analysis may still have a substantial probability of failure \citep{griffiths2004,cho2010}.

The price of this generality is computational. Each realisation requires a sequence of nonlinear finite-element analyses, and crude Monte Carlo requires on the order of $10^4$ to $10^5$ realisations to estimate a failure probability of $10^{-3}$. \citet{li2016a} reported 60,000 finite-element analyses and approximately 500 hours of computation for a direct Monte Carlo reference. Subset simulation \citep{au2001,li2016b} and importance sampling \citep{au2003} reduce the number of required samples, but each sample still requires a full elastoplastic analysis. Limit-equilibrium analyses of random fields are much cheaper \citep{cho2010}, but their failure probabilities can differ from finite-element estimates by a factor of about three \citep{li2016a}.

Surrogate models replace the expensive analysis with regression models trained on solver output. Gaussian-process regression on the Karhunen--Lo\`eve coefficients \citep{zhu2021}, neural networks and support-vector regression \citep{he2020}, gradient boosting \citep{wang2020}, failure classifiers \citep{aminpour2023}, Gaussian processes on reduced variables \citep{liu2024}, convolutional neural networks on the discretised random field \citep{zhang2026a} and graph neural networks on the finite-element mesh \citep{li2026} predict the factor of safety or the failure class of a realisation, and data-driven predictors of the factor of safety have also been used to shorten the SRM itself \citep{deng2026}. Image-to-image networks go further and predict the failure surface, the sliding volume and the deformation field of a spatially variable slope \citep{wang2024}. These studies leave three gaps. First, every training label comes from a conventional solver, and the number of solver runs ranges from a few hundred \citep{he2020} to $10^4$ \citep{li2026} and $10^5$ \citep{deng2026}, so the surrogate inherits the cost of the solver before it saves any, and no surrogate for slope reliability has been built from a source other than a finite-element or limit-equilibrium solver. Second, accuracy near the limit state requires training data in the failure region \citep{zhu2021}, several studies did not examine failure probabilities below $10^{-2}$ \citep{he2020,aminpour2023}, and whether a surrogate trained on Monte Carlo fields alone can be trusted at $\Pf\approx10^{-3}$ has not been tested against an independent estimate of the same probability. Third, with the exception of \citet{wang2024}, the surrogates return one scalar per realisation, so the probability that a zone lies in the slip band, the size of the sliding mass and the displacement response given failure, which all need the response of every sample, remain unavailable at the sample sizes that rare events demand.

Physics-informed neural networks (PINNs) \citep{raissi2019} remove the need for labels by training the network on the governing equations. In solid mechanics they have been formulated in strong form \citep{haghighat2021}, in energy and variational form \citep{samaniego2020,kharazmi2021,gao2022} and, most relevant here, on the finite-element mesh itself: the finite element-integrated neural network of \citet{zhang2025} evaluates the Galerkin weak form with shape functions, Gauss integration and an incremental stress update, and solves elastic and elastoplastic boundary-value problems without labelled data, with a modified Cam-clay model whose flow rule is associated. Energy-based losses extend to associated plasticity \citep{he2023}, but they require a potential and therefore do not apply to the non-associated flow rules used for soils. In geotechnical engineering, \citet{zhang2024} trained a PINN without labelled data to construct a kinematically admissible failure surface for the upper-bound limit analysis of homogeneous slopes, with an associated flow rule and without the stress field. \citet{liu2026} solved the elastoplastic Mohr--Coulomb displacement field of a homogeneous slope and of a slope with a weak interlayer with a label-free strong-form PINN that uses a non-associated flow rule and a return map detached from automatic differentiation, and obtained the factor of safety by slip-surface stress analysis of the stress field at the working strength, so the limit state is inferred from a single pre-failure stress state rather than reached by strength reduction. Other PINNs for Mohr--Coulomb plasticity still include labelled data in the loss \citep{yuan2026,zhang2026b}, including a Fourier-feature PINN by the present authors \citep{robjanghvad2026}, and for elastoplastic footings the forward analysis without data has been limited to elasticity, with plasticity treated as an inverse problem \citep{chen2025}. \citet{yuan2026}, \citet{chen2025} and \citet{robjanghvad2026} all report reduced accuracy in shear bands. Taken together, these studies leave the forward problem that the SRM poses unsolved by a PINN. The label-free formulations are limited either to an associated flow rule \citep{he2023,zhang2024,zhang2025} or to a single stress state at the working strength \citep{liu2026}, whereas the formulations that follow non-associated Mohr--Coulomb plasticity rely on finite-element labels \citep{yuan2026,zhang2026b,robjanghvad2026}. No PINN has carried the plastic history from step to step along an incremental strength-reduction path with a non-associated flow rule until a mechanism forms, none has been checked realisation by realisation against a finite-element SRM on random fields, and none has been run on the thousand or more realisations that a reliability analysis needs. A further obstacle is that a PINN solves one boundary-value problem at a time and must be retrained for every new configuration \citep{choi2026}, which is costly when $10^6$ random fields are needed.

Neural operators learn the map from an input function to the solution and evaluate new inputs at negligible cost \citep{lu2021,li2021,kovachki2024}. Operators that expand the solution in a reduced basis \citep{hesthaven2018,lanthaler2023,lu2022} are well suited to fields on a fixed mesh. Physics can be added to the operator loss \citep{wang2021}, but the operators applied in geotechnical engineering, for example to consolidation \citep{choi2026} and to the thermal regime of embankments \citep{xiao2025}, are trained on the output of conventional numerical solvers, and a recent assessment notes that the cost of generating that training set is often larger than the cost it saves \citep{kumar2026}. An operator for slope reliability whose training data contain no finite-element solution has not been reported, and no operator has been used to return the displacement path and the failure mode of a random-field slope rather than a scalar factor of safety.

A separate gap concerns the limit-equilibrium estimates against which these methods are usually judged. \citet{li2016a} found that limit-equilibrium and finite-element failure probabilities differ by a factor of about three, while \citet{cho2010} and \citet{zhu2021} found the critical surfaces of the two methods at similar positions. Whether the difference in $\Pf$ comes from the circular shape of the slip surface or from the limit-equilibrium formulation itself has not been separated on the same random fields, and how a bias of a few percent in $F$ propagates into $\Pf$ at $10^{-3}$ has not been quantified.

The research gap can therefore be stated in three parts. There is no label-free solver that reproduces the finite-element SRM for non-associated Mohr--Coulomb slopes with spatially variable strength, there is no reliability pipeline for such slopes that is free of finite-element solutions, and there is no estimate of the failure mechanism and the displacement response conditional on rare failure from such a pipeline. The objective of this study is to close these gaps by developing a PINN that performs incremental strength reduction without labelled data, training a neural operator on its solutions alone, and using both to estimate the failure probability, the failure mechanism and the displacement response of a spatially variable slope, with the finite-element SRM used only to verify the result and the heterogeneous Bishop method as the limit-equilibrium benchmark.

A natural question is why a neural solver is needed at all when the finite-element SRM already solves each realisation reliably. The answer is not speed per field, because on comparable hardware the PINN is not faster than Abaqus for a single slope (Section~\ref{sec:cost}), and this study does not claim otherwise. The benefit lies in four things that the finite-element workflow does not provide for a reliability analysis. First, the PINN is trained on the equations alone, so the chain from random field to failure probability needs no labelled data, no commercial solver and no licence per core, and the constitutive model is a few lines of differentiable code rather than a user subroutine. Second, because every realisation is an independent network on a shared mesh, hundreds of realisations are solved as one batched tensor program on a single GPU, whereas a finite-element code solves one realisation per job, so the cost of a Monte Carlo set scales with the number of batches rather than with the number of fields. Third, the PINN solutions are produced in the same framework as the operator that learns from them, so the hours spent on $10^3$ PINN solutions are paid once and buy an operator that evaluates $10^6$ fields in seconds, a sample size at which a direct finite-element Monte Carlo would take years (Section~\ref{sec:cost}). At the scale of this study the total cost is nevertheless comparable, about 11 GPU-hours of PINN runs for the two weak cohesions against an estimated 26 hours of sequential Abaqus runs for the same fields on the reference machine, so the argument is independence from a solver and from labelled data rather than a saving in cost. Fourth, once the operator is trained, the response is a smooth differentiable function of the KL coefficients, which gives the gradient that FORM needs by automatic differentiation and opens the way to inverse analysis against field measurements. The finite-element method keeps its role as the independent reference. It verifies the solver on three homogeneous slopes and 52 random fields, and that comparison is itself a result, because it shows what a label-free solver delivers and what it costs in accuracy.

The scope is a two-dimensional plane-strain $c$--$\varphi$ slope of one geometry, 3\,m high at 1:1, with cross-correlated lognormal cohesion and friction-angle fields of one anisotropic correlation structure, a Mohr--Coulomb model with zero dilation and no pore pressure, and three mean cohesions chosen so that the failure probability ranges from no failure in 500 fields to about $10^{-1}$, with a rare-event case near $10^{-3}$. The PINN--SRM is verified against Abaqus on the same mesh for three homogeneous slopes and 52 random fields, and the failure probability is estimated by crude Monte Carlo simulation, FORM on surrogates, importance sampling and the neural operator. Three-dimensional slopes, other geometries and correlation structures, hardening models and pore-water pressure are outside the scope and are discussed as future work. The contributions are as follows.
\begin{enumerate}
\item A label-free PINN performs incremental strength reduction on the finite-element mesh with a non-associated Mohr--Coulomb return map ($\psi=0$) and carries the plastic strain from one accepted strength-reduction step to the next. Because no potential exists for non-associated flow, the loss is the discrete virtual work with the stress held fixed, so that its gradient is the finite-element nodal residual, and the factor of safety follows from the same displacement-rate criterion as in the finite-element SRM.
\item The PINN--SRM is verified against Abaqus on three homogeneous slopes and on 52 individual random fields, with factors of safety within 2 to 3\,\% RMS and failure mechanisms in the same weak zones. The dependence of the result on the solver budget is quantified.
\item The failure probability is estimated from $10^{-1}$ to $10^{-3}$ by Monte Carlo simulation, FORM on surrogates and importance sampling. A heterogeneous Bishop analysis of the same fields underestimates $\Pf$ by factors of 2.6 and 8.3, and the critical circles show that the bias comes from the difference between the simplified Bishop formulation and the non-associated strength-reduction model rather than from the circular shape of the slip surface.
\item A reduced-basis neural operator trained only on the PINN solutions predicts the factor of safety, the displacement path and the failure mode of $10^6$ fields in seconds, so that no finite-element solution enters the reliability pipeline at any stage and Abaqus serves only to verify it. For $\Pf$ alone the operator is equivalent to a linear surrogate of $\ln F$, but it additionally yields the failure mechanism and the displacement response conditional on failure at $\Pf\approx10^{-3}$, which a scalar surrogate cannot provide.
\end{enumerate}
Section~\ref{sec:problem} defines the problem and the random fields. Section~\ref{sec:pinn} describes the PINN--SRM, and Section~\ref{sec:verif} verifies it against Abaqus. Section~\ref{sec:operator} presents the neural operator, and Section~\ref{sec:reliability} the reliability analysis with the results. Section~\ref{sec:discussion} discusses cost and limitations, and Section~\ref{sec:conclusions} concludes. Derivations and the mesh study are given in the appendices.

\section{Problem statement}\label{sec:problem}

\subsection{Geometry, loading and boundary conditions}
\begin{figure}[htbp]
\centering
\includegraphics[width=\textwidth]{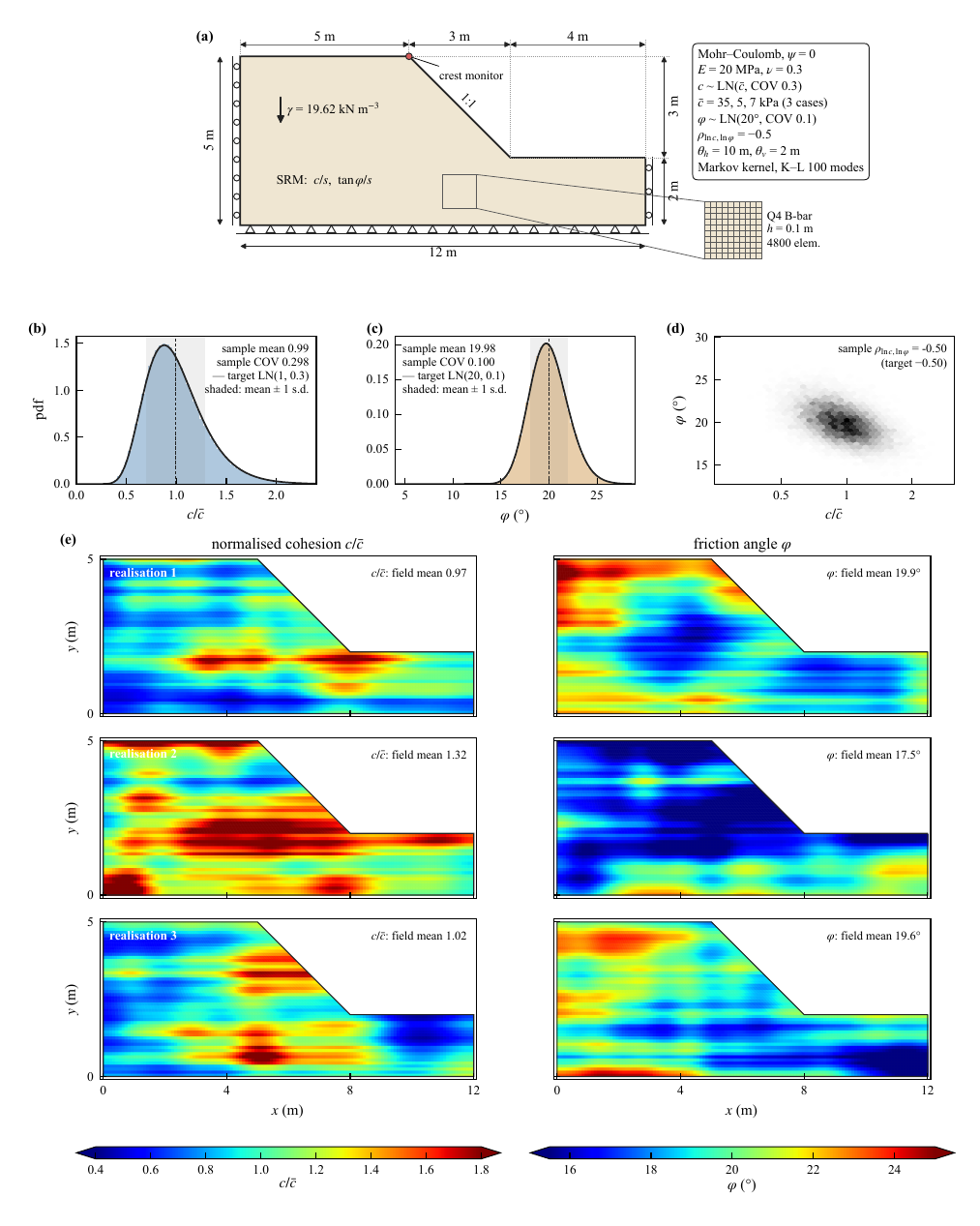}
\caption{Problem definition. (a) Slope geometry, boundary conditions and finite-element mesh (Q4 B-bar, $h=0.1$\,m, 4800 elements), common to the homogeneous benchmark ($c=\cb$, $\varphi=20^\circ$) and the random-field analyses. (b) Marginal distribution of $c/\cb$, (c) of $\varphi$, and (d) joint distribution of the Monte Carlo fields against the targets. (e) Three realisations of $c/\cb$ and $\varphi$.}
\label{fig:problem}
\end{figure}

The problem is a 2D plane-strain slope, as shown in Fig.~\ref{fig:problem}a. The domain is 12\,m wide and 5\,m high. The slope is 3\,m high with an inclination of 1:1 bounded by a 5\,m-wide crest and 4\,m-wide toe platform at a height of 2\,m. The base is fixed in both directions, while the two vertical sides are subject to roller constraints. The ground surface is traction-free. The only load is the self-weight of the soil, with $\gamma = 19.62$\,kN\,m$^{-3}$, and no pore pressure is considered. The settlement is monitored at the crest point $(x,y)=(5,5)$\,m. The domain is discretised using a structured mesh of 4800 bilinear quadrilateral elements ($h=0.1$\,m) that conforms to the slope face. The same mesh is used for the PINN and for Abaqus reference analyses (Section~\ref{sec:verif}). The sensitivity of the factor of safety to the mesh is examined in \ref{app:mesh}.

\subsection{Governing equations}\label{sec:governing}
\subsubsection{Equilibrium and boundary conditions}
Under quasi-static loading and small strains, the stress field $\bm\sigma$ in the domain $\Omega$ satisfies
\begin{equation}
\nabla\cdot\bm\sigma + \mathbf{b} = \mathbf{0}\quad\text{in }\Omega,\qquad \mathbf{b} = (0,-\gamma)^{\top},
\label{eq:equil}
\end{equation}
with the boundary conditions
\begin{equation}
\begin{aligned}
&\mathbf{u}=\mathbf{0}\ \text{on the base},\qquad u_x=0,\ \sigma_{xy}=0\ \text{on the lateral sides},\\
&\bm\sigma\cdot\mathbf{n}=\mathbf{0}\ \text{on the ground surface},
\end{aligned}
\label{eq:bc}
\end{equation}
where $\mathbf{u}=(u_x,u_y)$ is the displacement and $\mathbf{n}$ the outward normal.

\subsubsection{Kinematics and elasto-plasticity}
The strain is the symmetric displacement gradient and is split additively into an elastic and a plastic part,
\begin{equation}
\bm\varepsilon = \tfrac12\big(\nabla\mathbf{u}+\nabla\mathbf{u}^{\top}\big),\qquad \bm\varepsilon=\bm\varepsilon^{e}+\bm\varepsilon^{p},
\label{eq:strain}
\end{equation}
and the stress follows from the elastic strain through the isotropic linear elasticity tensor $\mathbb{D}$ ($E=20$\,MPa, $\nu=0.3$),
\begin{equation}
\bm\sigma = \mathbb{D}:\big(\bm\varepsilon-\bm\varepsilon^{p}\big).
\label{eq:hooke}
\end{equation}
The admissible stresses are bounded by the Mohr--Coulomb yield function in the form used by Abaqus,
\begin{equation}
\begin{aligned}
f(\bm\sigma;c,\varphi) &= R_{mc}(\Theta,\varphi)\,q - p\tan\varphi - c \le 0,\\
R_{mc} &= \frac{\sin(\Theta+\pi/3)}{\sqrt{3}\cos\varphi} + \frac13\cos(\Theta+\pi/3)\tan\varphi,
\end{aligned}
\label{eq:yield}
\end{equation}
where $p=-\tfrac13\operatorname{tr}\bm\sigma$ is the mean stress (compression positive), $q=\sqrt{3J_2}$ the deviatoric stress and $\Theta$ the deviatoric polar angle. The material is perfectly plastic. The plastic strain rate is normal to a plastic potential $g$,
\begin{equation}
\dot{\bm\varepsilon}^{p} = \dot\lambda\,\frac{\partial g}{\partial\bm\sigma},\qquad
g(\bm\sigma) = \sqrt{\big(\epsilon\,c\tan\psi\big)^2 + \big(R_{mw}\,q\big)^2} - p\tan\psi,
\label{eq:potential}
\end{equation}
which is the Menetrey--Willam surface \citep{menetrey1995} used by Abaqus, with the deviatoric shape $R_{mw}(\Theta,e)$ and meridional eccentricity $\epsilon=0.1$. The plastic multiplier satisfies the loading--unloading (Kuhn--Tucker) conditions
\begin{equation}
\dot\lambda\ge0,\qquad f\le0,\qquad \dot\lambda f=0.
\label{eq:kt}
\end{equation}
The dilation angle is zero, $\psi=0$, such that $g=R_{mw}\,q$ and the plastic flow is purely deviatoric. Zero dilation is commonly adopted in strength-reduction analyses and avoids the unrealistic volumetric expansion associated with an associated flow rule. Since $g\neq f$, the flow rule is non-associated, and the consistent tangent operator is generally nonsymmetric. Consequently, the elasto-plastic problem cannot be formulated as the minimisation of a potential energy. This non-associativity determines the choice of the loss function in Section~\ref{sec:pinn}.

\subsubsection{Weak form}
Multiplying Eq.~\eqref{eq:equil} by a kinematically admissible virtual displacement $\delta\mathbf{u}$ (zero where $\mathbf{u}$ is prescribed) and integrating by parts with the traction-free surface gives the principle of virtual work,
\begin{equation}
\int_\Omega \bm\sigma:\delta\bm\varepsilon\,\dd\Omega = \int_\Omega \mathbf{b}\cdot\delta\mathbf{u}\,\dd\Omega\qquad\forall\,\delta\mathbf{u},
\label{eq:vw}
\end{equation}
where $\bm\sigma$ is determined by Eqs.~\eqref{eq:hooke}--\eqref{eq:kt}. Equation~\eqref{eq:vw} is applicable to both associated and non-associated constitutive laws and forms the basis of the PINN loss (Section~\ref{sec:pinn}). Since the material response is path-dependent, Eqs.~\eqref{eq:hooke}--\eqref{eq:kt} are integrated incrementally along the loading path, which in this study is defined by the strength reduction process.

\subsection{Factor of safety by strength reduction}
The factor of safety (FoS) is evaluated with the strength reduction method (SRM). In the SRM, the soil strength parameters are reduced by a strength reduction factor $s$ \citep{potts2012}, such that
\begin{equation}
c_s = \frac{c}{s},\qquad \tan\varphi_s = \frac{\tan\varphi}{s},
\label{eq:srf}
\end{equation}
where the reduction is applied at each point in the domain. The strength reduction factor $s$ is increased until the slope can no longer be maintained in equilibrium with the reduced strength. The factor of safety $F$ is defined as the critical value of $s$ at which the prescribed failure criterion is reached. Because the material is path dependent, the stress state at each value of $s$ is obtained incrementally from the state at the previous value of $s$, with the plastic strain carried forward. The limit is identified with the same displacement-rate criterion in both the PINN and Abaqus (Section~\ref{sec:pinn}).

\subsection{Spatial variability of the strength}\label{sec:rf}
The cohesion $c(\mathbf{x})$ and the friction angle $\varphi(\mathbf{x})$ are lognormal random fields, as in earlier random-field slope analyses \citep{griffiths2004,cho2010}.
\begin{equation}
\ln c(\mathbf{x}) = \mu_{\ln c} + \sigma_{\ln c}\,g_c(\mathbf{x}),\qquad
\ln\varphi(\mathbf{x}) = \mu_{\ln\varphi} + \sigma_{\ln\varphi}\,g_\varphi(\mathbf{x}),
\label{eq:lognormal}
\end{equation}
with $\sigma_{\ln}^2=\ln(1+\mathrm{COV}^2)$ and $\mu_{\ln}=\ln(\text{mean})-\sigma_{\ln}^2/2$. The mean friction angle is $20^\circ$ with a COV of 0.1. The cohesion has a COV of 0.3 and a mean $\cb$ that defines the three cases below. $g_c$ and $g_\varphi$ are zero-mean, unit-variance Gaussian fields with the separable Markov correlation
\begin{equation}
\rho(\Delta x,\Delta y) = \exp\!\left(-\frac{2|\Delta x|}{\theta_h}-\frac{2|\Delta y|}{\theta_v}\right),\qquad \theta_h=10\text{ m},\ \theta_v=2\text{ m},
\label{eq:markov}
\end{equation}
and are cross-correlated through $g_\varphi=\rho_{c\varphi}\,g_c+\sqrt{1-\rho_{c\varphi}^2}\,g'$ with $\rho_{c\varphi}=-0.5$, where $g'$ is an independent field of the same kind. Each Gaussian field is represented by a truncated Karhunen--Lo\`eve (KL) expansion \citep{ghanem2003,phoon2002},
\begin{equation}
g(\mathbf{x}) \approx \sum_{k=1}^{M}\sqrt{\lambda_k}\,\xi_k\,\phi_k(\mathbf{x}),\qquad \xi_k\sim\mathcal{N}(0,1),
\label{eq:kl}
\end{equation}
where $(\lambda_k,\phi_k)$ are the eigenpairs of the correlation operator \eqref{eq:markov}, computed at the element centroids using area weights, and the truncated field is rescaled to have unit pointwise variance. With $M=100$ modes per field, each realisation is fully described by the standard-normal vector $\bm\xi\in\mathbb{R}^{200}$, and the element values of $c$ and $\varphi$ are evaluated at the element centroids. Fig.~\ref{fig:problem}b--d shows that the Monte Carlo fields reproduce the target marginal distributions and correlation structure, while Fig.~\ref{fig:problem}e shows three realisations. The discrete eigenproblem and the rescaling are given in \ref{app:kl}. The parameters are summarised in Table~\ref{tab:params}.

\begin{table}[htbp]
\centering
\caption{Parameters of the slope problem.}
\label{tab:params}
\begin{tabular}{ll}
\toprule
Parameter & Value \\
\midrule
Domain width $\times$ height & 12\,m $\times$ 5\,m \\
Slope height, inclination & 3\,m, 1:1 \\
Mesh & 4800 Q4 elements, $h=0.1$\,m \\
\addlinespace
Young's modulus, $E$ & 20\,MPa \\
Poisson's ratio, $\nu$ & 0.3 \\
Unit weight, $\gamma$ & 19.62\,kN\,m$^{-3}$ \\
Yield criterion & Mohr--Coulomb \\
Dilation angle, $\psi$ & $0^\circ$ \\
\addlinespace
Mean cohesion, $\cb$ & 35, 5 or 7\,kPa \\
COV of cohesion & 0.3 \\
Mean friction angle, $\bar\varphi$ & $20^\circ$ \\
COV of friction angle & 0.1 \\
Marginal distribution of $c$ and $\varphi$ & lognormal \\
Cross-correlation of the Gaussian fields, $\rho_{c\varphi}$ & $-0.5$ \\
Correlation function & separable Markov \\
Correlation lengths, $\theta_h$, $\theta_v$ & 10\,m, 2\,m \\
KL modes per field & 100 \\
Dimension of $\bm\xi$ & 200 \\
\bottomrule
\end{tabular}
\end{table}

\subsection{Cases and quantities of interest}\label{sec:cases}
Three mean cohesion values are considered with the same random-field model and the same set of standard-normal vectors $\bm\xi$ (fixed seed). Thus, the underlying spatial pattern is identical across cases, and the realisations differ only in the scale of the cohesion:
\begin{enumerate}
\item $\cb=35$\,kPa: a slope far from failure, in which the spatial variability changes the factor of safety and the shape of the slip surface, but failure is not expected.
\item $\cb=5$\,kPa: a marginally stable slope, with strength close to the failure limit and a high probability of failure.
\item $\cb=7$\,kPa: an intermediate slope in which failure is a rare event.
\end{enumerate}
The three cases thus span the range from a slope far from failure to a high probability of failure, with a rare-event case in between. The two weak cases were selected with a preliminary limit-equilibrium screening over $\cb$. For each case the homogeneous slope with $c=\cb$ and $\varphi=20^\circ$ everywhere serves as the deterministic reference and as the first verification problem. The quantities of interest are the distribution of $F$, the probability of failure
\begin{equation}
\Pf = P[F(\bm\xi)<1],
\label{eq:pf}
\end{equation}
the location of the slip surface and of the sliding mass, and the crest displacement accumulated along the strength-reduction path from $s_0$ to $s=1$, which is used in Section~\ref{sec:opmc} as a measure of the displacement response.

\section{Label-free PINN for incremental strength reduction}\label{sec:pinn}

\begin{figure}[p]
\centering
\includegraphics[width=\textwidth]{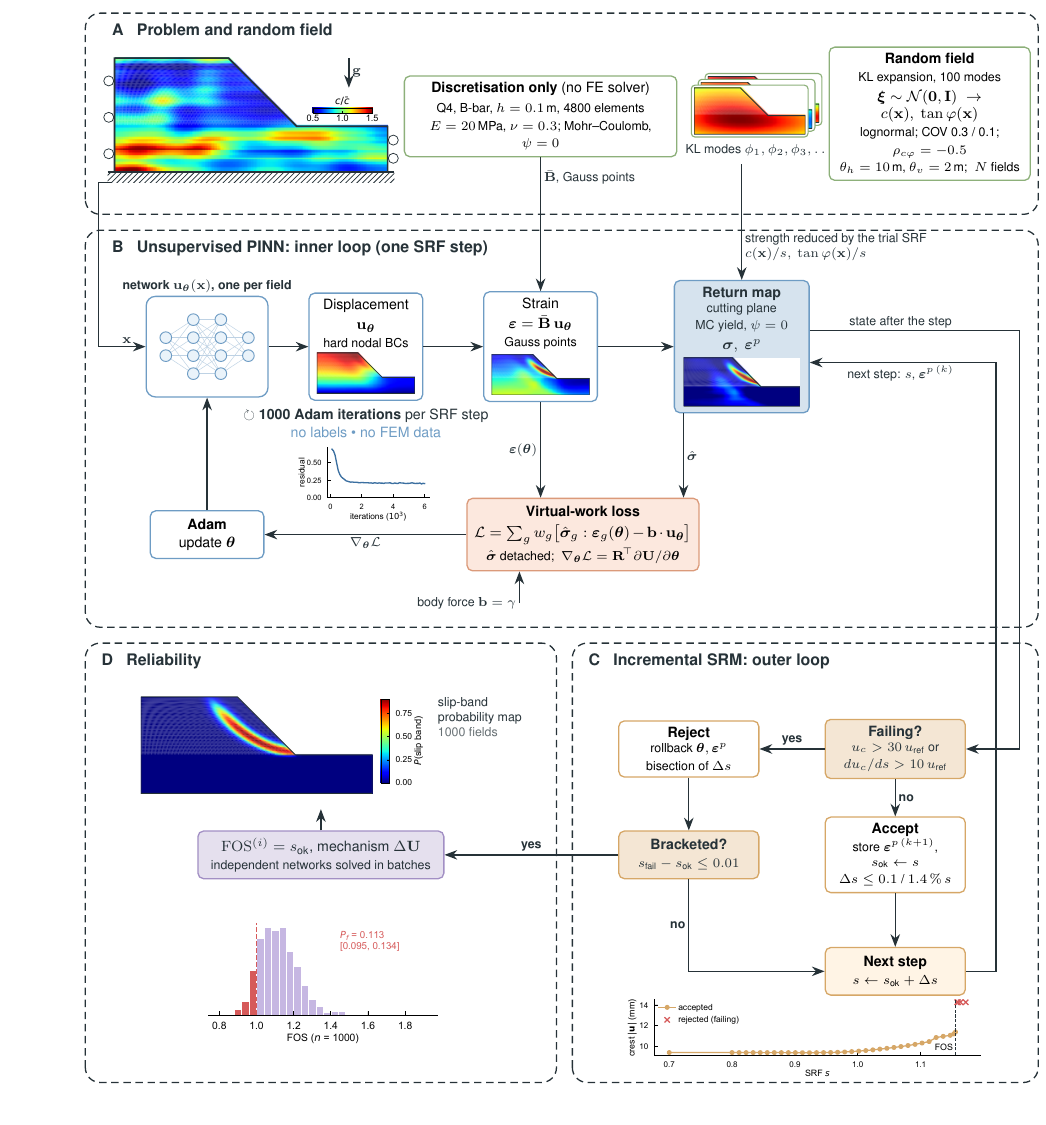}
\caption{Workflow of the label-free incremental PINN--SRM. (A) $c$--$\varphi$ realisations are generated by Karhunen--Lo\`eve expansion without FE data or an FE solver. (B) For each strength reduction factor $s$, the PINN is trained using a virtual-work loss, Mohr--Coulomb return mapping, and incremental plastic history. (C) Each step is accepted or rolled back, with bisection until the failure SRF is bracketed within 0.01. (D) The resulting factors of safety yield the failure probability and slip-surface probability map. Abaqus is used only for independent verification. The insets are computed results at $\cb=5$\,kPa: one random field with $F=0.97$ and its first three KL modes (A), the displacement, strain and plastic strain of that field at the last accepted SRF (B), the residual during the gravity step of the homogeneous slope (B), the SRF path with rejected trials of the homogeneous slope, run with the confirmation rule of \ref{app:budget} (C), and the slip-band probability and the distribution of $F$ over 1000 fields (D).}
\label{fig:workflow}
\end{figure}

We solve the strength-reduction problem described in Section~\ref{sec:problem} for each random field using a physics-informed neural network (PINN) \citep{raissi2019} trained only on the governing equations. The training is label-free in the sense that no labelled solution data enter the loss function: no displacement, stress, or factor-of-safety data from a finite-element solver are used. The workflow is summarised in Fig.~\ref{fig:workflow}.

\subsection{Displacement approximation and hard boundary conditions}
The network represents the nodal displacements of the mesh of Section~\ref{sec:problem}. For node $I$ with coordinates $\mathbf{x}_I$ it returns
\begin{equation}
u_I = U_0\,\frac{4x_I}{L_x}\Big(1-\frac{x_I}{L_x}\Big)\frac{y_I}{L_y}\,\mathcal{N}_u(\mathbf{x}_I;\bm\theta),\qquad
v_I = U_0\,\frac{y_I}{L_y}\,\mathcal{N}_v(\mathbf{x}_I;\bm\theta),
\label{eq:ansatz}
\end{equation}
where $\mathcal{N}=(\mathcal{N}_u,\mathcal{N}_v)$ is a fully connected network with parameters $\bm\theta$, $L_x=12$\,m, $L_y=5$\,m, and $U_0=0.01$\,m is a displacement scale. The displacement multipliers vanish on the base and, for $u$, on the two lateral boundaries, ensuring that the kinematic boundary conditions are satisfied exactly and do not enter the loss functions \citep{sukumar2022}. The network has five hidden layers of 128 tanh units. Its input at each node consists of the normalised coordinates, the local standardised values of $\ln c$ and $\ln\tan\varphi$ for the realisation, the KL vector $\bm\xi$ through a branch of width 128, and $\ln s$. Within each element, the displacement is interpolated from the nodal values using bilinear shape functions. Thus, the PINN and the displacement-based finite-element model share the same kinematics \citep{saha2021}. The strains are evaluated at the $2\times2$ Gauss points using an in-plane variant of the $\bar{B}$ \citep{hughes1980} treatment of the volumetric strain, which avoids volumetric locking under the isochoric plastic flow ($\psi=0$).

\subsection{Weak-form loss}\label{sec:loss}
As the flow rule is non-associated (Section~\ref{sec:governing}), energy-based losses, which minimise a potential as in elasticity or associated plasticity \citep{he2023,samaniego2020}, cannot be used. We instead train the PINN using the discrete form of the principle of virtual work in Eq.~\eqref{eq:vw} \citep{kharazmi2021}. The discretisation follows the finite element-integrated approach of \citet{zhang2025}, in which the network returns nodal values and the weak form is assembled with shape functions and Gauss integration, there for a modified Cam-clay model with an associated flow rule. Here the stress is detached from the computational graph, a device also used by \citet{liu2026} in a strong-form PINN, so that the loss remains defined for a non-associated flow rule without a potential. The stress $\bm\sigma_j$ at Gauss point $j$ is obtained from the constitutive update described in Section~\ref{sec:constitutive} and held fixed during differentiation (i.e., stop-gradient). The loss is
\begin{equation}
\mathcal{L}(\bm\theta) = \sum_j w_j\Big[\bm\sigma_j:\bm\varepsilon_j(\bm\theta) + \gamma\,v_j(\bm\theta)\Big],
\label{eq:loss}
\end{equation}
where $w_j$ denotes the Gauss weight multiplied by the Jacobian, $\bm\varepsilon_j$ is the $\bar{B}$ strain, and $v_j$ is the vertical displacement at the Gauss point.
\begin{equation}
\frac{\partial\mathcal{L}}{\partial\bm\theta} = \Big(\frac{\partial\mathbf{U}}{\partial\bm\theta}\Big)^{\!\top}\big(\mathbf{f}^{\mathrm{int}}(\mathbf{U})-\mathbf{f}^{\mathrm{ext}}\big) = \Big(\frac{\partial\mathbf{U}}{\partial\bm\theta}\Big)^{\!\top}\mathbf{R}(\mathbf{U}),
\label{eq:lossgrad}
\end{equation}
The loss gradient is the finite-element nodal residual $\mathbf{R}$ mapped to the network parameters. Taking $\delta\mathbf{u}$ as each nodal shape function in Eq.~\eqref{eq:vw} gives $\mathbf{R}=\mathbf{0}$. A stationary point of the training, $\partial\mathcal{L}/\partial\bm\theta=\mathbf{0}$, for a network capable of representing the nodal field therefore corresponds to an equilibrium state \eqref{eq:equil} of the discretised problem. The quality of the solution is monitored with the relative nodal residual $\|\mathbf{R}\|_{\mathrm{rms}}/\bar f$, where $\bar f=\gamma A/N_{\mathrm{nodes}}$ is the mean nodal gravity load. The derivation of Eq.~\eqref{eq:lossgrad} and the definition of the residual monitor are given in \ref{app:gradient}.

\subsection{Constitutive update and plastic history}\label{sec:constitutive}
At each Gauss point, the stress follows from the total strain $\bm\varepsilon$, the plastic strain $\bm\varepsilon^{p}_n$ from the last accepted SRF step, and the reduced strength $(c_s,\tan\varphi_s)$ defined in Eq.~\eqref{eq:srf}. Integrating Eqs.~\eqref{eq:hooke}--\eqref{eq:kt} over the step using a backward-Euler scheme gives the return map to the Mohr--Coulomb surface in Eq.~\eqref{eq:yield} \citep{simo1998}:
\begin{equation}
\begin{aligned}
&\bm\sigma = \mathbb{D}:\big(\bm\varepsilon-\bm\varepsilon^{p}_n-\Delta\bm\varepsilon^{p}\big),\qquad
\Delta\bm\varepsilon^{p} = \Delta\lambda\,\frac{\partial g}{\partial\bm\sigma},\\
&\Delta\lambda\ge0,\qquad f(\bm\sigma;c_s,\varphi_s)\le0,\qquad \Delta\lambda\,f=0.
\end{aligned}
\label{eq:returnmap}
\end{equation}
The update is solved using a cutting-plane algorithm with four iterations \citep{simo1985}, using closed-form gradients of $f$ and $g$ that were verified against automatic differentiation (\ref{app:constitutive}). The plastic strain $\bm\varepsilon^{p}_n$ is updated only when an SRF step is accepted. A rejected trial therefore never alters the history. Carrying this history is essential: solving each SRF from a zero plastic history (i.e., as a single-step analysis) gives a different, path-independent problem and a factor of safety that is too low.

\subsection{Incremental strength reduction}\label{sec:incremental}
Each realisation is analysed with one network that is carried along the SRF path \citep{griffiths1999}:
\begin{enumerate}
\item Gravity: The network, initialised at random, is trained for 6000 Adam iterations \citep{kingma2015} (learning rate $10^{-3}$) at a strengthened state $s_0=0.6$. Starting below $s=1$ ensures that the initial state is stable also for the weakest realisations. The maximum nodal displacement of this state, $u_{\mathrm{ref}}$, is the reference displacement of the realisation.
\item Trial steps: The SRF is increased by $\Delta s=0.1$ while the soil is elastic and by $\Delta s=0.014\,s$ once any Gauss point has yielded. Each trial step starts from the network of the last accepted state and runs 1000 Adam iterations with a fresh optimiser (learning rate $3\times10^{-4}$, reduced tenfold over the last 20\,\% of the step).
\item Acceptance: A trial with maximum nodal displacement $u$ is rejected when
\begin{equation}
u > 30\,u_{\mathrm{ref}}\qquad\text{or}\qquad \frac{u-u_{\mathrm{ok}}}{\Delta s} > 10\,u_{\mathrm{ref}},
\label{eq:failure}
\end{equation}
where $u_{\mathrm{ok}}$ is the displacement of the last accepted state. The test is a displacement-based form of the non-convergence criterion of the finite-element SRM \citep{dawson1999,griffiths1999}. A trial that diverges substantially is terminated before exhausting its iteration budget. An accepted trial updates $\bm\varepsilon^{p}_n$ and becomes the new starting point.
\item Bisection: After a rejection, the network is rolled back to the last accepted state and the step is halved, until the interval between the last accepted and the first rejected SRF is below 0.01. A rejection that occurred with a step larger than 0.025 is confirmed once with the final small step. The factor of safety is the last accepted SRF, $F=s_{\mathrm{ok}}$.
\end{enumerate}
The displacement increment between the first rejected trial and the last accepted state defines the failure mechanism of the realisation, from which the slip band (the 5\,\% of elements with the largest deviatoric strain increment) and the sliding mass (the elements whose displacement increment exceeds half of its maximum) are extracted. In the Abaqus reference analyses the same rate threshold, $10\,u_{\mathrm{ref}}$ per unit SRF, is applied to the crest displacement.

\subsection{Batched implementation}
Because the realisations are independent, a batch of $R$ networks (typically $R=100$) is trained simultaneously. The weights of all networks are stored as stacked tensors and evaluated using batched matrix products. Each network follows its own SRF schedule, with its own optimiser state, rollback, and bisection. One optimiser iteration for the entire batch is captured once as a CUDA graph \citep{paszke2019} and replayed, because the computational cost at this problem size is dominated by kernel launches rather than arithmetic. The acceptance test is evaluated every 50 iterations.

\section{Verification of the PINN--SRM against FEM (Abaqus)}\label{sec:verif}

\subsection{Homogeneous slopes}\label{sec:homog}
We first verify the solver on the three homogeneous slopes of Section~\ref{sec:cases} ($\cb=35$, 5, and 7\,kPa, $\varphi=20^\circ$), for which a conventional finite-element solution is well-established. The reference is Abaqus/Standard on the same mesh as the PINN (4800 CPE4 elements, whose selectively reduced integration is also a $\bar{B}$-type treatment of the volumetric strain, \ref{app:gradient}), with the same elastic constants, unit weight, and boundary conditions. The built-in Mohr--Coulomb model with $\psi=0$ and the default eccentricity of the flow potential is used, requiring the unsymmetric solver. Gravity is applied in a first step, and the strength is then reduced in a second step through a field variable that indexes tabulated values of $c/s$ and $\tan\varphi/s$, with increments of at most 0.004 of the step time. The factor of safety from Abaqus is obtained from the crest settlement using the same rate test as the PINN (Eq.~\eqref{eq:failure}). The last converged SRF, which is the usual non-convergence criterion, is reported as an upper bound. Applying gravity at SRF\,=\,1.0 instead of 0.6 changes the Abaqus factor of safety by less than 0.1\,\%. The simplified Bishop method with a circular slip search on a grid of centres and radii \citep{bishop1955} is included as a limit-equilibrium reference.

\begin{figure}[htbp]
\centering
\includegraphics[width=\textwidth]{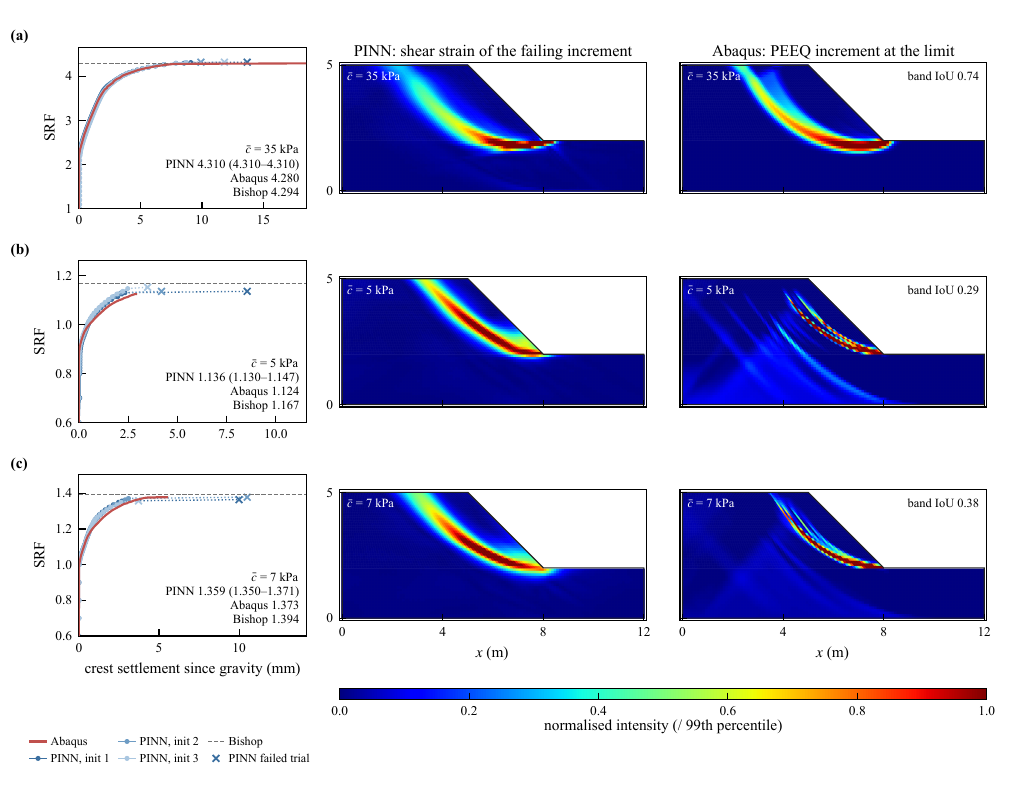}
\caption{Homogeneous slopes, PINN against Abaqus on the same mesh. (a)--(c) SRF against crest settlement for $\cb=35$, 5 and 7\,kPa: three PINN initialisations (crosses: failing trial), Abaqus and the Bishop factor of safety. Right: failure mechanism, shown as the shear-strain increment of the failing PINN trial and the PEEQ increment of Abaqus at the limit. IoU is the overlap of the thresholded shear bands.}
\label{fig:homog}
\end{figure}

Table~\ref{tab:homog} and Fig.~\ref{fig:homog} compare the factors of safety. The mean PINN value of three network initialisations lies within 1\,\% of Abaqus in all three cases ($F_{\mathrm{PINN}}/F_{\mathrm{Abaqus}}=1.007$, 1.010 and 0.990). The three initialisations give the same factor at $\cb=35$\,kPa and differ by 1.5\,\% at $\cb=5$ and 7\,kPa, where the limit is approached through a longer plastic path. Bishop is 0.3--3.8\,\% above Abaqus, the usual difference between limit equilibrium and strength reduction for a homogeneous slope \citep{griffiths1999}. The SRF--settlement curves of the PINN follow Abaqus up to the limit (Fig.~\ref{fig:homog}a--c). Both methods place the failure mechanism through the toe, deep for the strong slope and shallow for the weak ones (Fig.~\ref{fig:homog}, right). The overlap of the thresholded shear bands is 0.74 at $\cb=35$\,kPa but only 0.29 and 0.38 at $\cb=5$ and 7\,kPa. These bands are one to two elements thick, and the two measures are not identical (shear-strain increment of the failing trial for the PINN, PEEQ increment at the last converged state for Abaqus). Consequently, a shift of one element can halve the overlap, although the underlying mechanism is the same.

\begin{figure}[htbp]
\centering
\includegraphics[width=\textwidth]{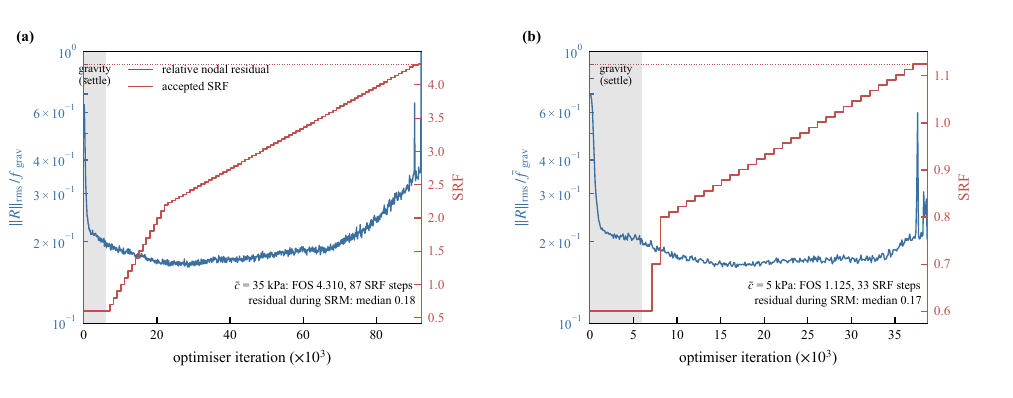}
\caption{Convergence of the PINN--SRM on the homogeneous slopes, (a) $\cb=35$\,kPa and (b) $\cb=5$\,kPa: RMS nodal residual normalised by the mean nodal gravity load (left axis, log scale) and accepted SRF (right axis) against optimiser iterations. The grey band is the gravity step, and the dotted line is the factor of safety.}
\label{fig:residual}
\end{figure}

We monitor the convergence of the solver through the FE nodal residual rather than through the value of the loss. The stress in Eq.~\eqref{eq:loss} is held fixed in the gradient, and the loss value therefore has no minimum at the solution and does not measure accuracy. The gradient of the loss, which equals the nodal residual of Eq.~\eqref{eq:lossgrad}, provides this measure. Fig.~\ref{fig:residual} shows the RMS nodal residual, normalised by the mean nodal gravity load, for $\cb=35$ and 5\,kPa. During the gravity step the residual falls from about 0.65 to 0.20. Along the strength reduction the residual remains on a plateau (median 0.18 and 0.17, 5--95\,\% range 0.16--0.31 and 0.16--0.21) and changes little within each SRF step. The 1000 iterations of a step mainly carry the state to the new SRF rather than reduce the residual further. The residual therefore does not vanish: a level of about 0.17 is the accuracy floor of the present network and optimiser, and the field errors reported below quantify the effect of this floor on the displacement and stress. Near the limit the residual rises (to about 0.35 at $\cb=35$\,kPa) as the plastic zone spreads. In the failing trial the residual increases by a factor of two or more, together with the displacement that triggers the failure test.

\begin{figure}[htbp]
\centering
\includegraphics[width=\textwidth]{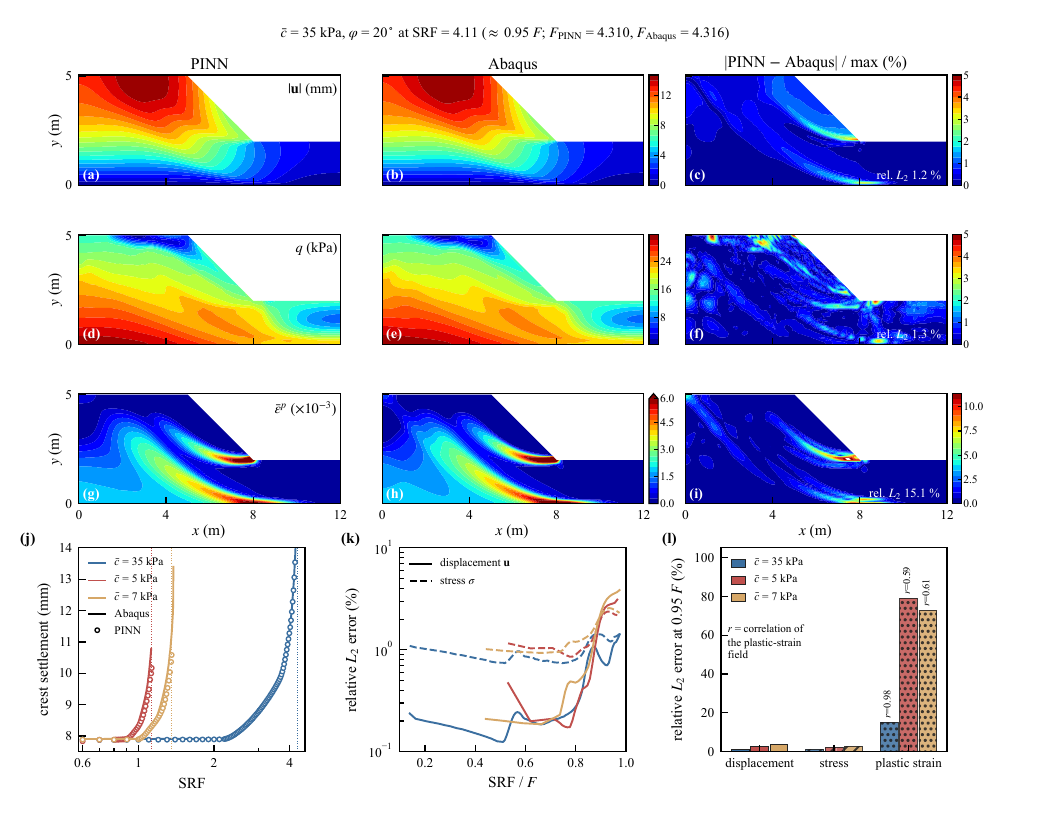}
\caption{Homogeneous slope, PINN against Abaqus on the same mesh. (a)--(i) $\cb=35$\,kPa at SRF $\approx0.95F$: displacement magnitude, deviatoric stress $q$ and equivalent plastic strain (rows), with PINN, Abaqus and their difference normalised by the Abaqus maximum (columns). (j) Crest settlement against SRF for $\cb=35$, 5 and 7\,kPa. (k) Relative $L_2$ error of displacement and stress along the SRF path. (l) Relative $L_2$ errors at $0.95F$ and correlation $r$ of the plastic-strain field.}
\label{fig:fields}
\end{figure}

Fig.~\ref{fig:fields} compares the internal fields along the whole SRF path. For the strong slope at $0.95F$ the relative $L_2$ errors are 1.2\,\% for the displacement and 1.3\,\% for the stress, and the equivalent plastic strain has the same pattern as Abaqus (Fig.~\ref{fig:fields}a--i), with a correlation of 0.98 and an $L_2$ error of 15\,\%. Up to $0.8F$ the displacement and stress errors remain near 1\,\% in all three cases and increase to 2--4\,\% at $0.95F$, when the plastic zone spreads (Fig.~\ref{fig:fields}k), and the crest settlement at that level is 0.9, 3.8 and 4.9\,\% below Abaqus for $\cb=35$, 5 and 7\,kPa (Fig.~\ref{fig:fields}j). The plastic strain of the weak slopes is the least accurate quantity. The location agrees with Abaqus, but the magnitude at $0.95F$ is about one quarter of the Abaqus value (correlations of 0.59 and 0.61, $L_2$ error 79 and 73\,\%, Fig.~\ref{fig:fields}l), because the budget of 1000 iterations per plastic step does not allow the plastic flow to converge fully within each step. The factor of safety is governed by the stress state at the limit rather than by the accumulated plastic strain and is nevertheless reproduced. The plastic-strain magnitude for weak slopes is interpreted qualitatively. With this procedure fixed, the PINN--SRM is applied to random fields in Section~\ref{sec:rfverif}.

\begin{table}[htbp]
\centering
\caption{Factor of safety of the homogeneous slopes ($\varphi=20^\circ$). PINN: mean of three network initialisations.}
\label{tab:homog}
\begin{tabular}{lcccc}
\toprule
$\cb$ (kPa) & PINN & Abaqus & Bishop & PINN / Abaqus \\
\midrule
35 & 4.310 & 4.280 & 4.294 & 1.007 \\
5 & 1.136 & 1.124 & 1.167 & 1.010 \\
7 & 1.359 & 1.373 & 1.394 & 0.990 \\
\bottomrule
\end{tabular}
\end{table}

\subsection{Random-field slopes}\label{sec:rfverif}
The homogeneous check does not test the feature that motivates the method: a failure mechanism that follows the weak zones of a heterogeneous field. The PINN--SRM is compared with Abaqus on 52 individual random fields, each run by both solvers on the same mesh with the same element values of $c$ and $\varphi$. The Abaqus models use one material per element, with $c/s$ and $\tan\varphi/s$ tabulated against the field variable, and otherwise follow Section~\ref{sec:homog}. At $\cb=35$\,kPa, 25 fields span the distribution of the factor of safety, including the lowest and highest values. At $\cb=5$\,kPa, 24 fields were selected based on the Bishop factor of safety before any PINN run: 11 from the lower tail, 11 at selected quantiles of the distribution and 2 from the upper end. These are complemented by three fields that lie on the limit state $F\approx1$: the most probable failure fields (FORM design points) identified using three surrogates of $F$, whose construction is described in Section~\ref{sec:reliability}. Because a small error in $F$ can move such a field across the limit state, these represent the most demanding cases for a failure/safe decision.

\begin{figure}[htbp]
\centering
\includegraphics[width=\textwidth]{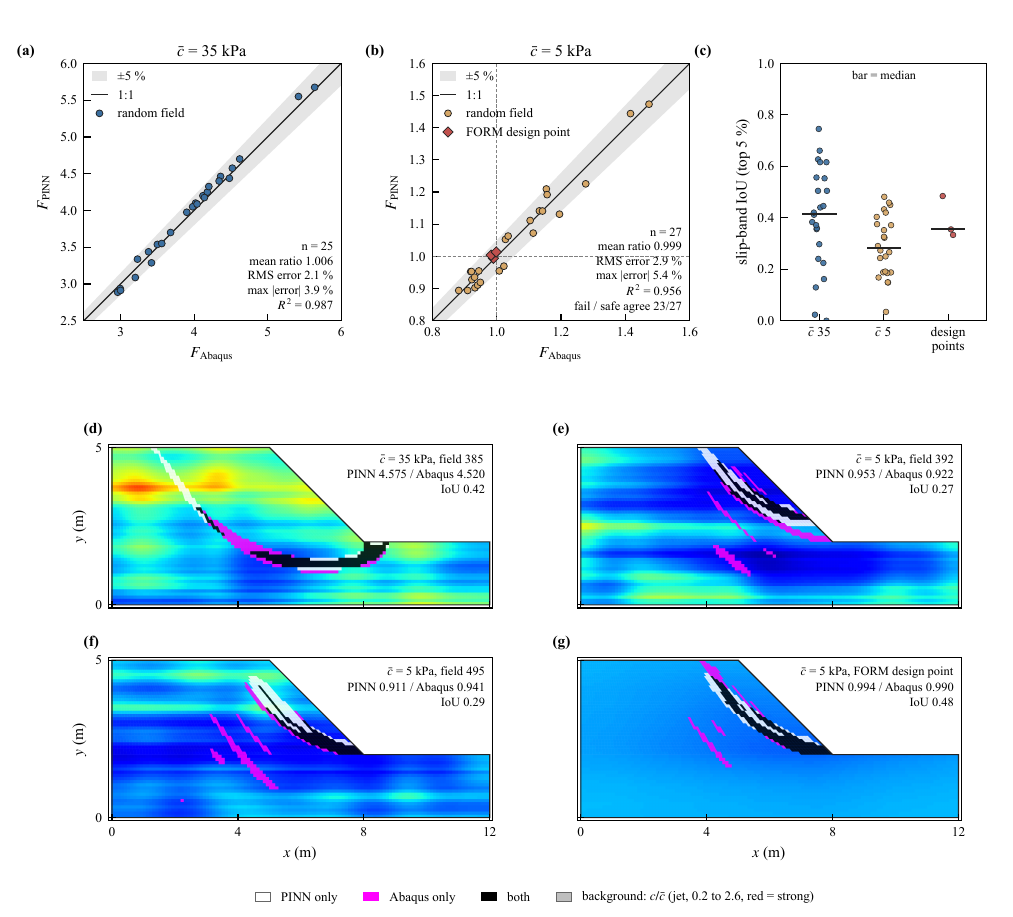}
\caption{PINN against Abaqus on 52 random fields (same mesh, same element properties). (a) $\cb=35$\,kPa, 25 fields. (b) $\cb=5$\,kPa, 24 fields and three FORM design points. The dashed lines mark $F=1$. The grey band is $\pm5$\,\%. (c) Overlap (IoU) of the slip bands, defined as the most strained 5\,\% of the elements in each solver, and the bar is the median. (d)--(g) Slip bands of fields with median overlap and of the linear-surrogate design point, over the cohesion field $c/\cb$ (jet, red = strong).}
\label{fig:rf}
\end{figure}

Fig.~\ref{fig:rf}a,b compares the factors of safety field by field. At $\cb=35$\,kPa ($F$ from 3.0 to 5.6), the mean ratio $F_{\mathrm{PINN}}/F_{\mathrm{Abaqus}}$ is 1.006, with an RMS error of 2.1\,\%, a maximum error of 3.9\,\%, and $R^2=0.987$. At $\cb=5$\,kPa ($F$ from 0.88 to 1.47, including the design points), the mean ratio is 0.999 with an RMS error of 2.9\,\%. In total, 24 of the 27 fields are within 5\,\%, and the maximum error is 5.4\,\%.

The errors have no systematic sign and contribute scatter rather than bias to the distribution of $F$. Of the 27 weak-soil fields, 23 are classified identically as failing ($F<1$) or safe by the two solvers. The four disagreements, comprising two random fields and two design points, all have $F_{\mathrm{Abaqus}}$ between 0.98 and 1.03 and lie within 5.3\,\% of the reference. These misclassifications are consistent with the approximately 3\,\% scatter near the limit state. For comparison, the heterogeneous Bishop analysis of the same fields is 3.6\,\% above Abaqus on average at $\cb=5$\,kPa, with a maximum difference of 8.2\,\%. It classifies only 10 of the 13 fields that fail in Abaqus as failing. This one-sided error is consistent with the lower Bishop failure probabilities reported in Section~\ref{sec:comparison}.

The failure mechanisms are compared through the most strained 5\,\% of the elements in each solver (Fig.~\ref{fig:rf}c--g). The median overlap of the two bands is 0.42 at $\cb=35$\,kPa, 0.28 at $\cb=5$\,kPa and 0.35 for the design points. These values are of the same order as for the homogeneous slopes, where the mechanism is known to be the same. The fields in Fig.~\ref{fig:rf}d--g were selected at the median overlap rather than at the best. In these fields the main bands coincide over most of their length and exit at the same point of the toe. The overlap is reduced by two effects rather than by a different mechanism. In the weak soil, Abaqus develops secondary parallel bands behind the main one (red in Fig.~\ref{fig:rf}e--g), which enter the Abaqus top 5\,\% and are absent from the single failing increment of the PINN. In the deep mechanism of the strong soil (Fig.~\ref{fig:rf}d), the strain of Abaqus concentrates in the lower part of the band, and the upper part, which the PINN resolves, falls below the 5\,\% threshold. Complete disagreement (overlap below 0.05) occurs in three fields (two at $\cb=35$\,kPa, one at $\cb=5$\,kPa). The factors of safety of these fields still differ by only 0.3--4.8\,\%, which indicates competing mechanisms of almost equal strength.

In summary, over 52 random fields the PINN--SRM reproduces the factor of safety of a commercial finite-element code to about 2--3\,\% RMS without bias and places the failure mechanism in the same weak zone. This accuracy is the basis for using the PINN solutions as training data for the neural operator in Section~\ref{sec:operator}.

\section{Neural operator trained on the PINN solutions}\label{sec:operator}

\begin{figure}[p]
\centering
\includegraphics[width=\textwidth]{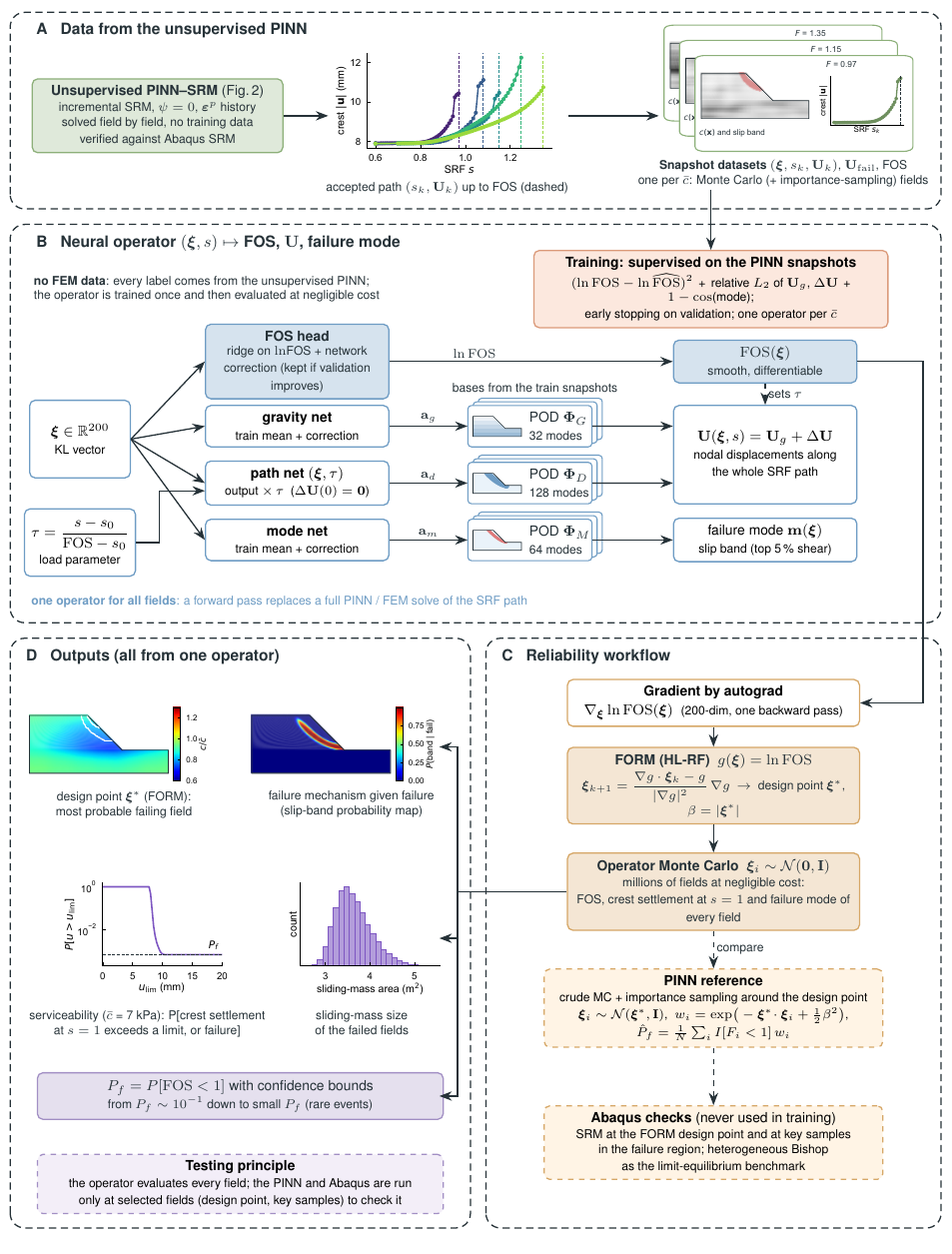}
\caption{Neural operator trained on label-free PINN solutions. (A) The incremental PINN--SRM of Fig.~\ref{fig:workflow} supplies accepted SRF paths from Monte Carlo fields and, for small failure probabilities, importance-sampling fields concentrated in the failure region. No FEM data are used. (B) The operator maps the KL vector $\bm\xi$ and normalised load $\tau$ to the factor of safety, displacement path represented by POD bases, and failure mode. (C) Autograd identifies the FORM design point, while millions of operator evaluations provide Monte Carlo estimates. (D) Outputs include the most probable failing field, the failure mechanism and sliding mass conditional on failure, exceedance of the crest displacement along the strength-reduction path, and $\Pf$ for rare events.}
\label{fig:operator}
\end{figure}

The computational cost of a PINN solution is about 15 to 20\,s per realisation (Section~\ref{sec:cost}), which is adequate for $10^3$ fields but not for the $10^6$ or more fields required for a direct Monte Carlo estimate of a small $\Pf$. The PINN solutions are used instead to train a neural operator, following the operator-learning framework of \citet{kovachki2024,lu2021}, that maps the random field, represented by the KL vector $\bm\xi$ of Eq.~\eqref{eq:kl}, directly to the response of the slope (Fig.~\ref{fig:operator}). The operator is trained only on PINN solutions. No finite-element result enters the training, and the Abaqus analyses of Section~\ref{sec:verif} serve only as independent checks. The physics enters through the PINN solutions, which satisfy the weak form of Eq.~\eqref{eq:vw}, whereas the training loss of the operator itself contains no physics term.

\subsection{Training data}
Each PINN realisation supplies the KL vector $\bm\xi\in\mathbb{R}^{200}$, the factor of safety $F$, the nodal displacement $\mathbf{U}_k$ at every accepted SRF $s_k$ along the path $s_0<s_1<\dots<s_K=F$, and the displacement increment of the failing trial. Three quantities are formed from these outputs. The first is the displacement after the gravity step, $\mathbf{U}_g=\mathbf{U}_0$. The second is the displacement increment along the strength-reduction path, $\Delta\mathbf{U}_k=\mathbf{U}_k-\mathbf{U}_0$, indexed by the normalised strength reduction
\begin{equation}
\tau_k = \frac{s_k-s_0}{F-s_0}\in[0,1].
\label{eq:tau}
\end{equation}
The third is the failure mode $\mathbf{m}$, the displacement increment of the failing trial normalised to unit length. The normalisation by $F$ places the limit state of every realisation at $\tau=1$.

At $\cb=5$\,kPa, the data comprise the 1000 Monte Carlo realisations of the PINN, split into 800 training, 100 validation, and 100 test fields. The 24 fields validated with Abaqus in Section~\ref{sec:rfverif} are all included in the test set. At $\cb=7$\,kPa, failure is rare, with only one of the 1000 Monte Carlo realisations failing. The 300 importance-sampling realisations from Section~\ref{sec:reliability}, which are concentrated in the failure region (128 of them fail), are therefore added to the data set, giving 1020 training, 130 validation, and 150 test fields. The fields checked with Abaqus are again included in the test set. The two data sets contain approximately $4\times10^4$ and $6\times10^4$ SRF states, respectively.

\subsection{Proper orthogonal decomposition (POD) of the displacement fields}
In this section, a nodal displacement field contains about $10^4$ degrees of freedom. Instead of predicting the displacement at every node, each field quantity is represented using a small set of spatial modes obtained by proper orthogonal decomposition (POD) \citep{chatterjee2000,sirovich1987}. POD is the data-driven counterpart of the KL expansion of Section~\ref{sec:rf}: KL decomposes the covariance of the input field into orthogonal modes, and POD decomposes the collected output fields in the same way, with the modes computed from the solutions rather than from an assumed covariance. The snapshots of a quantity from the training fields are stacked as the rows of a matrix $\mathbf{X}$, and the singular value decomposition $\mathbf{X}=\mathbf{W}\bm\Sigma\bm\Phi^{\top}$ gives the orthonormal modes $\bm\Phi$, ordered by the energy $\Sigma_{ii}^2$ that they capture. The snapshots are not centred, and the Euclidean inner product of the nodal vectors is used. A field is then represented by the coefficients of the first $r$ modes,
\begin{equation}
\mathbf{U}_g\approx U_s\,\bm\Phi_G\,\mathbf{a}_g,\qquad
\Delta\mathbf{U}(\tau)\approx U_s\,\bm\Phi_D\,\mathbf{a}_d(\tau),\qquad
\mathbf{m}\approx\bm\Phi_M\,\mathbf{a}_m,
\label{eq:pod}
\end{equation}
with $r=32$, 128 and 64 modes for the gravity state, the path increment and the failure mode, and a displacement scale $U_s$ equal to the mean norm of $\mathbf{U}_g$. Twelve states per training field, evenly spaced along the path, build the path basis. The bases are computed from the training fields only. The projection error of Eq.~\eqref{eq:pod} on the test fields, which is the lower bound of the operator error, has a median of 0.05\,\% for the gravity state, 0.6\,\% for the path increment near the limit ($\tau>0.9$) and about 1\,\% for the failure mode, at both mean cohesions. A network that predicts POD coefficients from the input parameters is a reduced-basis neural operator, in the family of the POD-NN \citep{hesthaven2018}, PCA-Net \citep{lanthaler2023} and POD-DeepONet approaches \citep{lu2022}. The operator of the present study differs mainly in the source of the training data, which are label-free PINN solutions rather than finite-element results.

\subsection{Operator heads}
The operator consists of four heads that share the input $\bm\xi$. Each head is written as a fixed base predictor plus a network correction whose last layer is initialised to zero,
\begin{equation}
\hat{\mathbf{y}}(\bm\xi) = \mathbf{y}_{\mathrm{base}}(\bm\xi) + \bm\sigma_y\odot\mathcal{N}(\bm\xi),
\label{eq:head}
\end{equation}
where $\bm\sigma_y$ is a fixed scale (the standard deviation of the target over the training fields, or 0.05 for $\ln F$) and $\mathcal{N}$ is a multilayer perceptron with SiLU activations \citep{elfwing2018}. Training therefore starts from the best simple predictor, and the network learns only the remaining dependence on $\bm\xi$. The correction is kept only when the validation loss decreases.
\begin{enumerate}
\item Factor of safety. The base is a ridge regression of $\ln F$ on $\bm\xi$ \citep{hoerl1970}, with the regularisation weight selected by cross-validation, and the correction is an ensemble of five networks (two hidden layers of 64). The correction lowers the validation error by less than 1\,\% at $\cb=5$\,kPa and not at all at $\cb=7$\,kPa, and the factor-of-safety head is in effect a linear model of $\ln F$ in $\bm\xi$. The logarithm keeps $F$ positive and makes the dependence on the lognormal strength closer to linear.
\item Gravity state. The base is the mean of $\mathbf{a}_g$ over the training fields. The correction has two hidden layers of 128. The loss is the squared error relative to $\|\mathbf{a}_g\|^2$.
\item Displacement path. The input is $(\bm\xi,\tau)$, extended by $\tau^2$ and $\sqrt\tau$, and the output is
\begin{equation}
\mathbf{a}_d(\bm\xi,\tau) = \tau\left[\bar{\mathbf{a}}_d(\tau) + \bm\sigma_d\odot\mathcal{N}_d(\bm\xi,\tau)\right],
\label{eq:pathhead}
\end{equation}
where $\bar{\mathbf{a}}_d(\tau)$ is the mean path of the training fields, interpolated in $\tau$, and $\mathcal{N}_d$ has four hidden layers of 384. The factor $\tau$ enforces $\Delta\mathbf{U}(0)=\mathbf{0}$. Each state is weighted by the inverse squared norm of the final increment of the realisation, which gives stiff and soft fields the same weight.
\item Failure mode. The base is the mean of $\mathbf{a}_m$. The correction has three hidden layers of 256. Because the mode has unit length, the loss is $1-\cos(\hat{\mathbf{a}}_m,\mathbf{a}_m)$.
\end{enumerate}
The networks are trained with AdamW \citep{loshchilov2019} and a cosine learning-rate schedule \citep{loshchilov2017}. The weights with the lowest validation loss are kept. Training takes about one minute per mean cohesion. Further convergence results are given in Section~\ref{sec:optraining}.

For a new field, the operator returns $F$ from the first head and the complete displacement path by evaluating Eq.~\eqref{eq:pod} at any $\tau$, with $s=s_0+\tau(F-s_0)$. The slip band and the sliding mass follow from the predicted failure mode with the same definitions as for the PINN (Section~\ref{sec:incremental}). One evaluation for $10^3$ fields, including 40 states of the displacement path, takes about 10\,ms. The operator is also differentiable with respect to $\bm\xi$, which gives the gradient of $F$ required by FORM (Section~\ref{sec:reliability}) by automatic differentiation \citep{paszke2019}.

\subsection{Training and accuracy}\label{sec:optraining}
\begin{figure}[htbp]
\centering
\includegraphics[width=\textwidth]{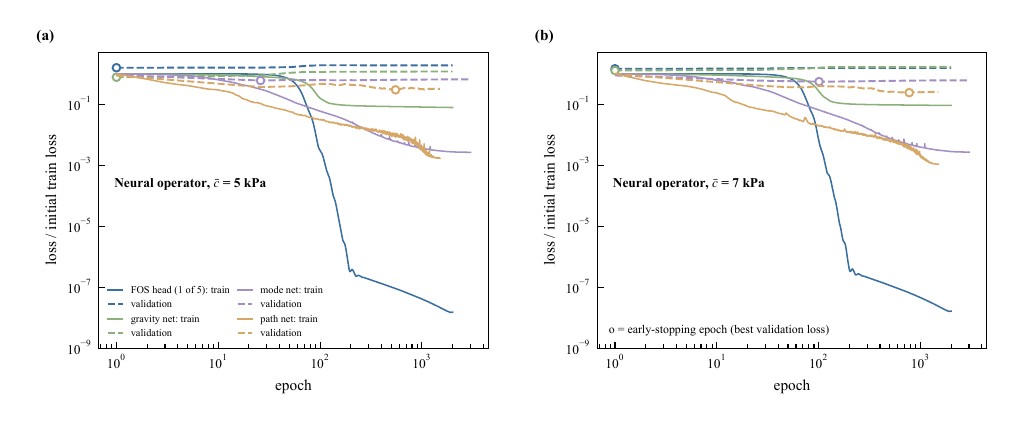}
\caption{Training of the neural operator at (a) $\cb=5$\,kPa and (b) $\cb=7$\,kPa: training and validation losses of each head, normalised by the initial training loss. Circles mark the early-stopping epoch.}
\label{fig:optrain}
\end{figure}

Fig.~\ref{fig:optrain} shows the training and validation losses of the four heads. The training losses decrease by two to eight orders of magnitude, whereas the validation losses decrease much less and reach their minima early. As noted above, correcting the factor-of-safety head provides almost no improvement. For the gravity state, the lowest validation loss occurs at the first epoch, and the prediction is effectively the mean gravity state of the training fields. The gravity state depends little on the strength field because the elastic properties are uniform and the plastic zone at $s_0$ is small. The mean already reproduces the gravity state of the validation fields with a relative squared error of $4\times10^{-6}$ to $6\times10^{-6}$. The corrections for the failure mode and displacement path are retained at epochs 25 and 550 for $\cb=5$\,kPa, and at epochs 100 and 775 for $\cb=7$\,kPa. They reduce the validation loss by about one quarter for the failure mode and by about two thirds for the path. The large gap between the training and validation losses indicates that 800 to 1000 fields provide only weak constraints for a network with 200 inputs. Early stopping prevents the corrections from fitting this scatter.

\begin{figure}[htbp]
\centering
\includegraphics[width=\textwidth]{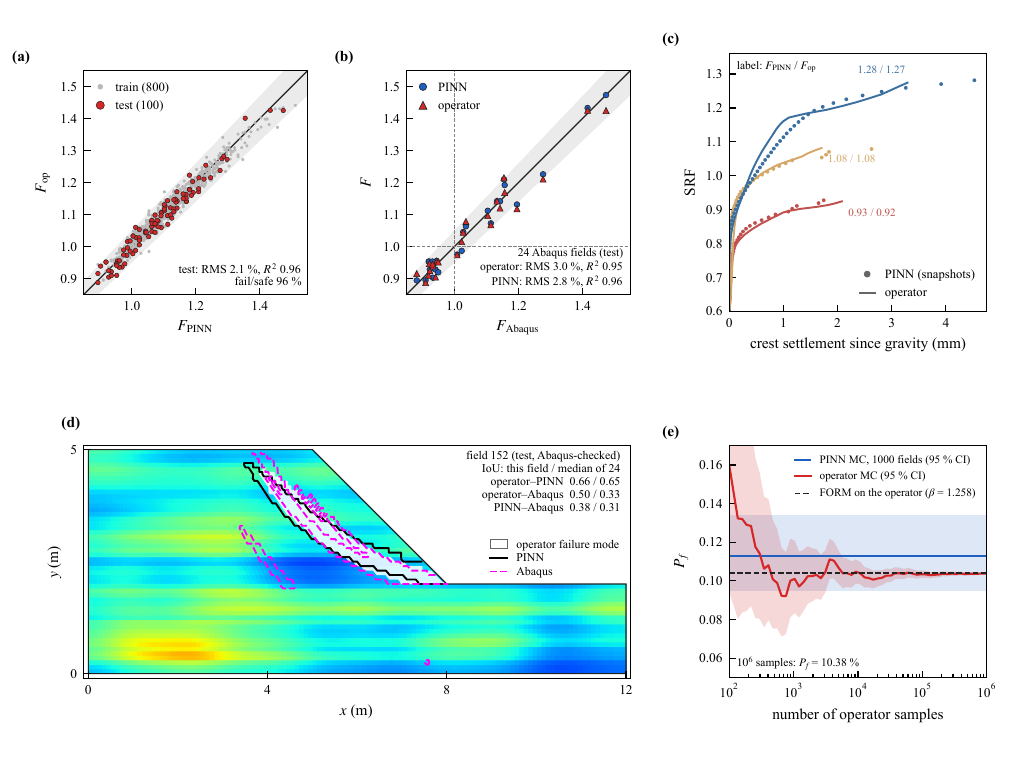}
\caption{Neural operator at $\cb=5$\,kPa. (a) $F$ against the PINN on training and test fields. (b) PINN and operator against Abaqus. (c) SRF--settlement paths of three test fields. (d) Failure mode of one Abaqus-checked field. (e) Convergence of the operator Monte Carlo estimate of $\Pf$.}
\label{fig:op5}
\end{figure}

Fig.~\ref{fig:op5} compares the operator with the PINN and with Abaqus at $\cb=5$\,kPa. On the 100 test fields the factor of safety predicted by the operator has an RMS error of 2.1\,\% with respect to the PINN ($R^2=0.96$), and 96\,\% of the fields are classified identically as failing or safe (Fig.~\ref{fig:op5}a). On the 24 test fields evaluated with Abaqus, the RMS error of the operator is 3.0\,\% ($R^2=0.95$), compared with 2.8\,\% ($R^2=0.96$) for the PINN (Fig.~\ref{fig:op5}b). The operator therefore introduces little additional error relative to the PINN on which it is trained. The predicted SRF--settlement paths follow the PINN states up to failure (Fig.~\ref{fig:op5}c). The crest settlement at $s=1$ on the held-out fields has an RMS error of 0.16\,mm at $\cb=5$\,kPa and 0.11\,mm at $\cb=7$\,kPa, for settlements of 8 to 10\,mm (Section~\ref{sec:opmc}). The predicted failure mode occupies the same band as the PINN mode (Fig.~\ref{fig:op5}d). Over the 24 Abaqus fields, the median overlap of the slip bands is 0.65 between the operator and the PINN, 0.33 between the operator and Abaqus and 0.31 between the PINN and Abaqus. The operator therefore agrees with Abaqus to a similar extent as the PINN does.

\section{Reliability analysis}\label{sec:reliability}

We compute the failure probability defined by Eq.~\eqref{eq:pf} using four comparative approaches, which differ in their analytical assumptions and the number of PINN computations required. All approaches, together with a heterogeneous Bishop limit-equilibrium model used as a baseline, are evaluated using the same set of standard-normal vectors $\bm\xi$ described in Section~\ref{sec:cases}. Individual formulations and their corresponding predictive performance are presented in the subsequent subsections.

\subsection{Heterogeneous Bishop analysis}
The simplified Bishop method described in Section~\ref{sec:homog} is extended to the random fields \citep{bishop1955,cho2010}. The slope is divided into 240 vertical slices, and trial circles are generated on a grid of $40\times40$ centres with 40 radii considered for each centre. The cohesion and friction angle at the base of each slice are taken from the element nearest to the midpoint of the base, and the factor of safety for each field is the minimum over all admissible circles. The analysis differs from the PINN in two respects: the slip surface is restricted to a circle, and limit equilibrium replaces strength reduction of an elastoplastic continuum. The comparison therefore captures the combined effect of both simplifications on the failure probability.

\subsection{Crude Monte Carlo and lognormal fit}\label{sec:mc}
For $N$ independent fields $\bm\xi_i\sim\mathcal{N}(\mathbf{0},\mathbf{I})$, the crude Monte Carlo estimate is \citep{griffiths2004}
\begin{equation}
\hat P_f = \frac1N\sum_{i=1}^{N} I[F(\bm\xi_i)<1],
\label{eq:mc}
\end{equation}
where $I[\cdot]$ equals 1 when the condition holds and 0 otherwise. The PINN is run on $N=500$ fields at $\cb=35$\,kPa and $N=1000$ fields at $\cb=5$ and 7\,kPa. The 95\,\% interval of $\hat P_f$ is the Wilson score interval \citep{wilson1927}, which remains valid when few or no failures are observed. The coefficient of variation of $\hat P_f$ is $\sqrt{(1-\Pf)/(N\Pf)}$, and a 10\,\% coefficient of variation at $\Pf=10^{-3}$ would require about $10^5$ fields. The PINN is run on the Monte Carlo fields for all three cases, whereas only at $\cb=5$\,kPa does the sample contain enough failures to provide an accurate estimate. At $\cb=35$\,kPa no field fails, and at $\cb=7$\,kPa only one field fails. A lognormal fit of the Monte Carlo sample provides a second, parametric estimate,
\begin{equation}
P_f^{\mathrm{LN}} = \Phi\!\left(-\frac{m_{\ln F}}{s_{\ln F}}\right),
\label{eq:lnfit}
\end{equation}
where $m_{\ln F}$ and $s_{\ln F}$ are the sample mean and standard deviation of $\ln F$ and $\Phi$ is the standard normal distribution function. The fit extrapolates into the tail from the bulk of the sample and is therefore accurate only when the lower tail of $F$ is lognormal.

\begin{figure}[htbp]
\centering
\includegraphics[width=\textwidth]{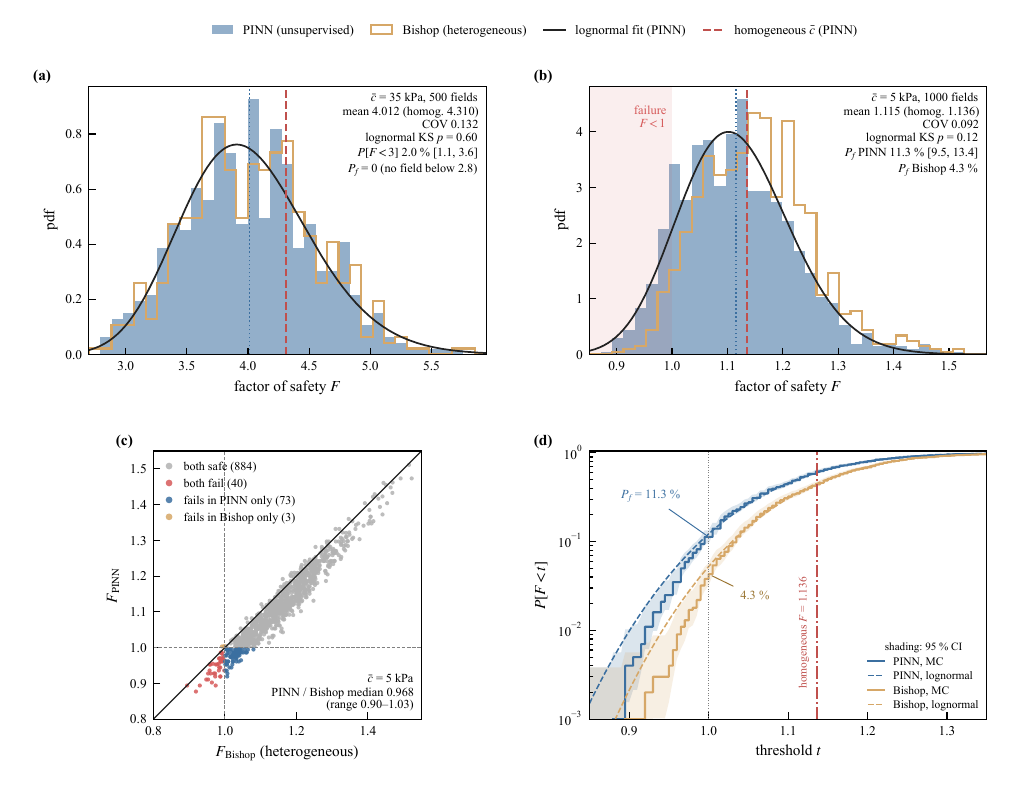}
\caption{Monte Carlo distribution of the factor of safety. (a) $\cb=35$\,kPa, 500 fields. (b) $\cb=5$\,kPa, 1000 fields: PINN, heterogeneous Bishop, lognormal fit and homogeneous $F$. (c) PINN against Bishop per field. (d) $P[F<t]$ with 95\,\% intervals. The vertical line is the deterministic (homogeneous) answer, for which $P[F<t]$ jumps from 0 to 1 at $t=F$.}
\label{fig:mc}
\end{figure}

The resulting distributions are illustrated in Fig.~\ref{fig:mc}. At $\cb=35$\,kPa the mean factor of safety of the 500 fields is 4.01 with a coefficient of variation of 0.13, which is 7\,\% below the homogeneous value of 4.31. The spatial variability therefore lowers the mean factor of safety in addition to spreading it. No field fails (the lowest $F$ is 2.84), and the Wilson upper bound of $\Pf$ is $7.6\times10^{-3}$. At $\cb=5$\,kPa the mean is 1.115 with a coefficient of variation of 0.092, 2\,\% below the homogeneous value of 1.136. Of the 1000 fields, 113 fail, and $\hat P_f=0.113$ with a 95\,\% interval of $[0.095,0.134]$. The homogeneous analysis ($F=1.136$) would declare the same slope safe. The Kolmogorov--Smirnov test does not reject the lognormal model \citep{massey1951} ($p=0.60$ and 0.12), although the test is insensitive to the tail, as Section~\ref{sec:is} shows, and at $\cb=5$\,kPa, Eq.~\eqref{eq:lnfit} gives $P_f^{\mathrm{LN}}=0.122$, within the Monte Carlo interval.

The heterogeneous Bishop analysis gives an almost identical mean factor of safety at $\cb=35$\,kPa (4.03), where the failure mechanism is deep and close to circular. At $\cb=5$\,kPa the Bishop factor of safety is higher (mean 1.152), and the median ratio of PINN to Bishop is 0.968 with a range of 0.90 to 1.03 (Fig.~\ref{fig:mc}c). Of the 1000 fields, 40 fail in both analyses, 73 fail only in the PINN and 3 fail only in Bishop. The failure probability estimated via Bishop's method is 0.043 $[0.032,0.057]$, which is 2.6 times lower than the corresponding PINN prediction. The factors of safety of the two analyses are nonetheless strongly correlated ($r=0.97$ at both $\cb=5$ and 7\,kPa). \citet{li2016a} reported a weaker correlation ($\rho=0.80$) between a limit equilibrium analysis restricted to a single slip surface and a finite-element analysis of an undrained slope. The full circle search of the present Bishop analysis follows the variation of the mechanism more closely, and the difference between the two analyses is mainly a shift of $F$ rather than scatter.

\begin{figure}[htbp]
\centering
\includegraphics[width=\textwidth]{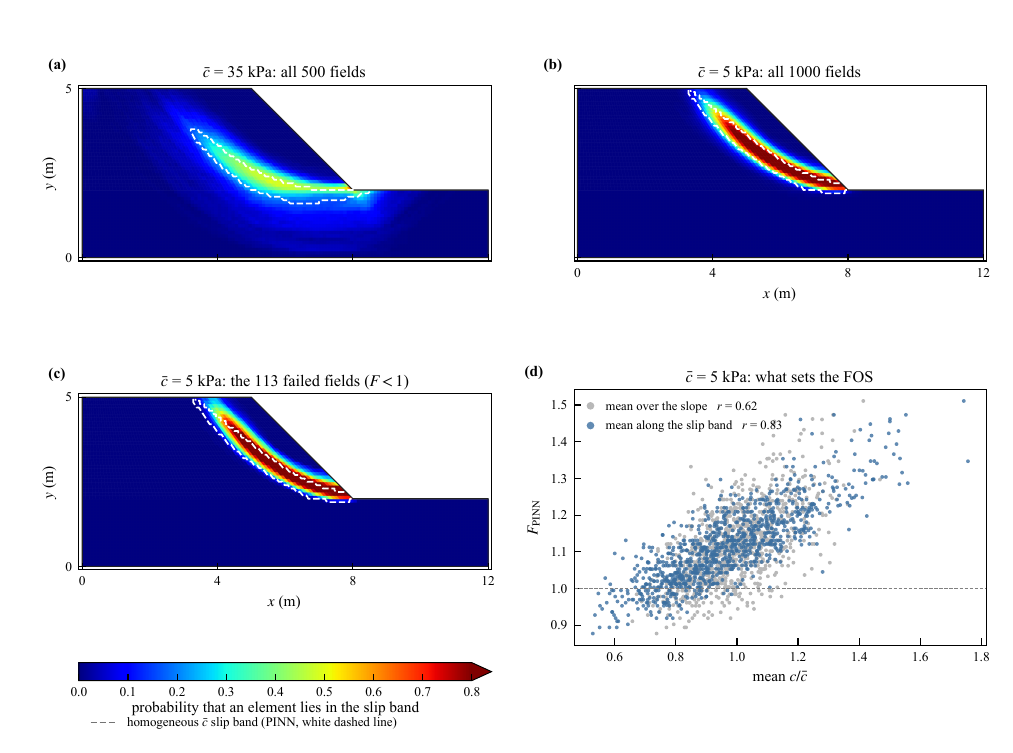}
\caption{Slip-band probability from the PINN solutions, i.e. the fraction of fields in which an element lies in the slip band. (a) $\cb=35$\,kPa, 500 fields. (b) $\cb=5$\,kPa, 1000 fields. (c) $\cb=5$\,kPa, the 113 failing fields. The dashed line is the slip band of the homogeneous slope. (d) $F$ against the mean of $c/\cb$ over the slope and along the slip band.}
\label{fig:slipband}
\end{figure}

The PINN solutions also provide the probability that an element lies in the slip band (Fig.~\ref{fig:slipband}). At $\cb=35$\,kPa the mechanism is deep and varies from field to field. The band probability reaches only 0.61 and spreads well beyond the band of the homogeneous slope. A similar departure of the critical slip surfaces of random fields from the deterministic surface was observed in finite-element analyses of undrained slopes \citep{li2016a}. At $\cb=5$\,kPa the mechanism is shallow and runs along the slope face, with a maximum band probability of 0.91 over all fields and 0.97 over the 113 failing fields. The factor of safety correlates more strongly with the mean cohesion along the slip band ($r=0.83$) than with the mean cohesion of the whole slope ($r=0.62$) (Fig.~\ref{fig:slipband}d). The weak zones crossed by the mechanism, rather than the average strength, control the factor of safety.

\subsection{Surrogates of the factor of safety and FORM}\label{sec:form}
The first-order reliability method (FORM) \citep{hasofer1974} locates the design point $\bm\xi^*$, the point of the limit state $g(\bm\xi)=\ln F(\bm\xi)=0$ closest to the origin of the standard-normal space. Because the density of $\bm\xi$ decreases with $\|\bm\xi\|$, the design point is the most probable failure field \citep{low2007}. The reliability index is $\beta=\|\bm\xi^*\|$, and the FORM estimate is $P_f^{\mathrm{FORM}}=\Phi(-\beta)$. The search requires the gradient $\nabla g$, which the PINN--SRM does not provide directly. We replace the limit state by a surrogate of $\ln F(\bm\xi)$ fitted to the PINN solutions of the first 500 Monte Carlo fields at $\cb=5$\,kPa. Three surrogates of increasing flexibility are used:
\begin{enumerate}
\item a linear model $\ln F\approx a+\mathbf{w}^{\top}\bm\xi$, fitted by ridge regression \citep{hoerl1970} with the regularisation weight chosen by cross-validation,
\item a Gaussian process with the sum of a linear and a squared-exponential kernel and a noise term, with the hyperparameters fitted by maximum likelihood,
\item an ensemble of eight multilayer perceptrons (two hidden layers of 64, tanh activations), each trained on a bootstrap resample of the fields, with the ensemble mean as the prediction.
\end{enumerate}
The accuracy of each surrogate is measured by five-fold cross-validation on the 500 fields. The design point is found with the Hasofer--Lind--Rackwitz--Fiessler (HL-RF) iteration \citep{rackwitz1978,hasofer1974},
\begin{equation}
\bm\xi_{k+1} = \frac{\nabla g(\bm\xi_k)^{\top}\bm\xi_k - g(\bm\xi_k)}{\|\nabla g(\bm\xi_k)\|^2}\,\nabla g(\bm\xi_k),\qquad \bm\xi_0=\mathbf{0},
\label{eq:hlrf}
\end{equation}
stopped when $\|\bm\xi_{k+1}-\bm\xi_k\|<10^{-6}$. The gradient is exact for the linear model, computed by finite differences for the Gaussian process and computed by automatic differentiation for the ensemble. For the linear model the iteration converges in one step to the closed form
\begin{equation}
\bm\xi^* = -\frac{a\,\mathbf{w}}{\|\mathbf{w}\|^2},\qquad \beta=\frac{a}{\|\mathbf{w}\|}.
\label{eq:linearform}
\end{equation}

\begin{figure}[htbp]
\centering
\includegraphics[width=\textwidth]{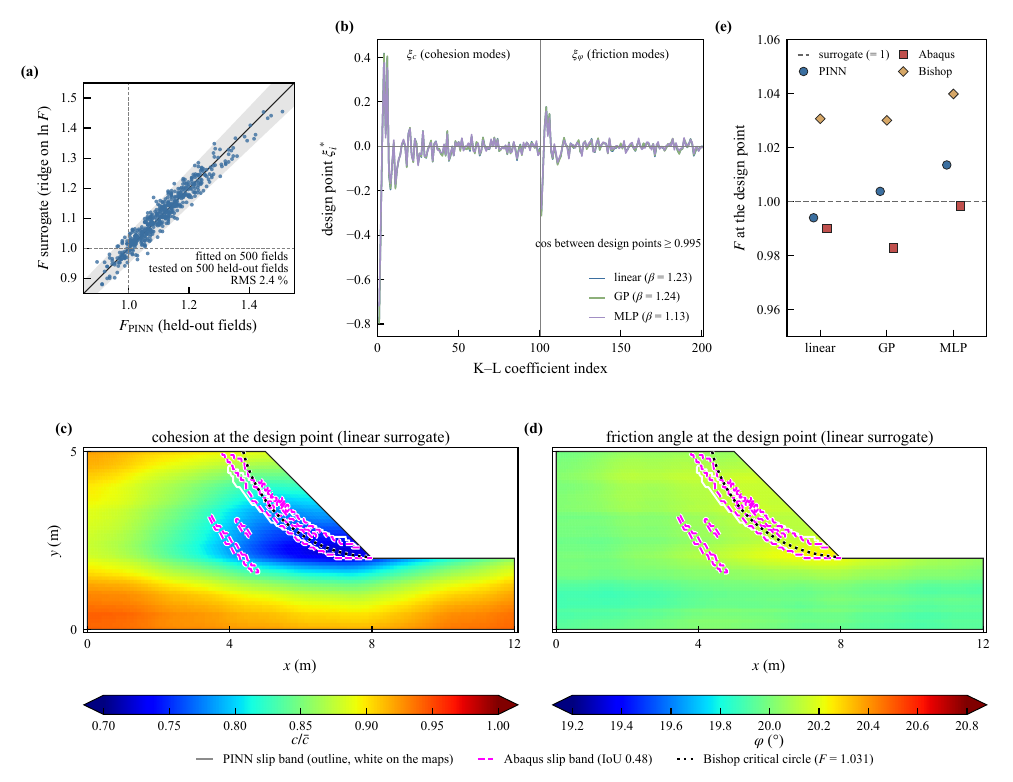}
\caption{FORM design point from surrogates of the PINN at $\cb=5$\,kPa. (a) Linear surrogate against the PINN on 500 fields not used in the fit. (b) Design points of the three surrogates in the KL coordinates. (c) Cohesion and (d) friction angle at the design point of the linear surrogate, with the PINN and Abaqus slip bands and the critical Bishop circle. (e) $F$ at the three design points from the PINN, Abaqus and Bishop. The surrogate value is 1 by construction.}
\label{fig:form}
\end{figure}

The three design points are fields that the PINN has not seen during training. They are evaluated independently with the PINN and Abaqus to verify whether each surrogate identifies the limit state consistently with both solvers. The three surrogates have cross-validated RMS errors of 2.6, 2.6 and 2.9\,\% and classify 96, 96 and 94\,\% of the fields correctly. The linear surrogate, tested on the other 500 Monte Carlo fields, has an RMS error of 2.4\,\% and a maximum error of 8.0\,\% (Fig.~\ref{fig:form}a). The added flexibility of the Gaussian process and of the ensemble does not improve the accuracy, which indicates that $\ln F$ is close to linear in $\bm\xi$ over the range of the data. The reliability indices are 1.23, 1.24 and 1.13, which give $P_f^{\mathrm{FORM}}=0.109$, 0.107 and 0.130, all inside the Monte Carlo interval. The three design points are almost the same field, with pairwise cosines of at least 0.995 (Fig.~\ref{fig:form}b). The cohesion modes carry 86\,\% of $\|\bm\xi^*\|^2$, and the first five modes carry 78\,\%. At the design point the cohesion is reduced to 0.86 of the mean on average and to 0.72 in a weak zone below the slope face, whereas the friction angle stays close to $20^\circ$ (Fig.~\ref{fig:form}c,d). The PINN gives $F=0.994$, 1.004 and 1.014 at the three design points and Abaqus gives 0.990, 0.983 and 0.998. Both solvers therefore confirm that the design points lie on the limit state within 2\,\%. The Bishop analysis gives 1.03 to 1.04 at the same fields (Fig.~\ref{fig:form}e). At the design point of the linear surrogate, the PINN and Abaqus slip bands pass through the same weak zone (overlap 0.48). The critical Bishop circle of this field runs inside the same band (Fig.~\ref{fig:form}c,d), in agreement with \citet{zhu2021}, who found the critical surfaces of a random limit equilibrium analysis and of random finite-element analyses at similar positions in a single-layer slope. Nevertheless, the Bishop factor of safety is 1.031, about 4\,\% above both solvers. At this field the difference therefore comes from the formulation, the simplified Bishop method against a non-associated strength-reduction model, rather than from the shape of the slip surface. The difference is of the same order as for the homogeneous slopes (Section~\ref{sec:homog}).

\subsection{Importance sampling for the rare-event case}\label{sec:is}
\begin{figure}[htbp]
\centering
\includegraphics[width=\textwidth]{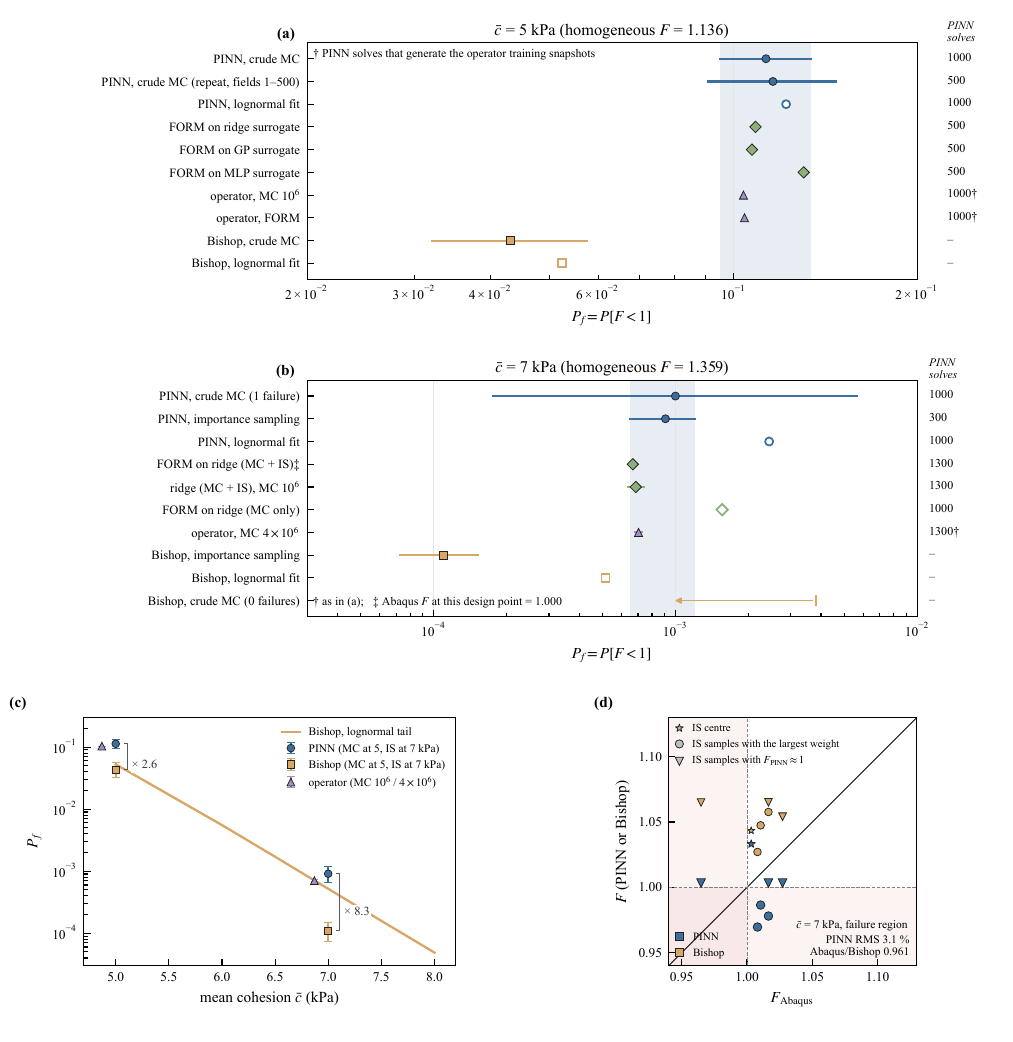}
\caption{Probability of failure from each estimator. (a) $\cb=5$\,kPa, (b) $\cb=7$\,kPa, with the number of PINN solutions each needs. The shaded band is the 95\,\% interval of the PINN reference (crude Monte Carlo in (a), importance sampling in (b)). Bars show 95\,\% intervals, and estimators without a bar provide no interval. Open markers are extrapolations into the tail from the bulk of the sample or from Monte Carlo fields only. The arrow is an upper bound (no failing field). (c) $\Pf$ against the mean cohesion, with the ratio of the PINN to the Bishop estimate. The operator markers are shifted to the left for clarity. (d) $\cb=7$\,kPa: PINN and Bishop against Abaqus on fields of the failure region. Colour denotes the method and shape the group of fields. The shading marks $F<1$.}
\label{fig:estimators}
\end{figure}

When $\cb=7$\,kPa, only one failure occurs among the 1,000 Monte Carlo realizations, giving a crude estimate of $\hat P_f=1.0\times10^{-3}$. The corresponding confidence interval, $[1.8\times10^{-4},5.6\times10^{-3}]$, is wide and spans more than one order of magnitude. A more precise reference value of $\Pf$ is obtained from the PINN using importance sampling \citep{au2003} with the shifted density $h(\bm\xi)=\varphi_{200}(\bm\xi-\bm\xi_c)$, where $\varphi_{200}$ denotes the standard-normal density in $\mathbb{R}^{200}$ centred at $\bm\xi_c$. A density of this form, centred near a design point with unit covariance, remains efficient in high dimensions \citep{au2003}. Using $N_{\mathrm{IS}}$ samples $\bm\xi_i\sim h$, the estimate is
\begin{equation}
\hat P_f^{\mathrm{IS}} = \frac{1}{N_{\mathrm{IS}}}\sum_{i=1}^{N_{\mathrm{IS}}} I[F(\bm\xi_i)<1]\,w_i,\qquad
w_i = \frac{\varphi_{200}(\bm\xi_i)}{\varphi_{200}(\bm\xi_i-\bm\xi_c)} = \exp\!\big(-\bm\xi_c^{\top}\bm\xi_i+\tfrac12\|\bm\xi_c\|^2\big).
\label{eq:is}
\end{equation}
The estimate is unbiased for any centre, and the centre affects only the variance. The centre must be selected before the PINN runs, and the design point of the PINN is not yet known at that stage. We take as the centre a proxy design point obtained from the heterogeneous Bishop analysis of the 1000 Monte Carlo fields: the Bishop factors of safety are scaled by 0.968, the median ratio of PINN to Bishop at $\cb=5$\,kPa, and Eq.~\eqref{eq:linearform} is applied to a ridge fit of their logarithm, which gives $\|\bm\xi_c\|=3.03$. At $\cb=5$\,kPa, where the PINN design points are available, the same construction yields a centre whose direction differs little from that of the PINN design points (cosine of 0.97). The use of a cheap limit equilibrium analysis to guide the sampling of a more expensive model follows the idea of \citet{li2016a}, who combined subset simulation with a limit equilibrium analysis to select the samples for finite-element analyses. In both cases the estimate remains that of the expensive model, because the limit equilibrium analysis only directs the sampling. The PINN is run on $N_{\mathrm{IS}}=300$ samples, and the 95\,\% interval of Eq.~\eqref{eq:is} is the percentile interval of a bootstrap over the samples \citep{efron1993}. The same samples are evaluated with the heterogeneous Bishop analysis, which gives the Bishop estimate with the same estimator. The 300 samples also provide the operator with training fields in the failure region (Section~\ref{sec:operator}). The linear surrogate and FORM are then applied to all 1300 PINN solutions (Monte Carlo and importance sampling), and the surrogate is also sampled directly with $10^6$ fields. A second linear surrogate fitted to the 1000 Monte Carlo fields alone shows how far a surrogate trained without failure-region data extrapolates into the tail. Of the 300 importance samples, 128 fail, and $\hat P_f^{\mathrm{IS}}=9.1\times10^{-4}$ with a 95\,\% interval of $[6.5,12.0]\times10^{-4}$ and a coefficient of variation of 15\,\% (Table~\ref{tab:pf7}). The lognormal fit of the Monte Carlo sample gives $2.4\times10^{-3}$, 2.7 times the importance-sampling value. The lower tail of $F$ at $\cb=7$\,kPa is therefore thinner than lognormal, and a fit to the bulk of the sample overestimates the rare-event probability. The linear surrogate fitted to all 1300 solutions gives $\beta=3.21$ and $P_f^{\mathrm{FORM}}=6.7\times10^{-4}$, and direct sampling of the same surrogate gives $6.9\times10^{-4}$. Abaqus gives $F=1.000$ at the design point of this surrogate. The surrogate fitted to the Monte Carlo fields alone gives $\beta=2.96$ and $1.6\times10^{-3}$, a factor of 1.7 above the importance-sampling value. Failure-region data are therefore needed for an accurate tail. \citet{zhu2021} reached the same conclusion for a Gaussian-process surrogate of a $c$--$\varphi$ slope with $\rho_{c\varphi}=-0.5$: the surrogate trained on randomly sampled fields was inaccurate near $F=1$ until training points in the failure region were added. The Bishop analysis of the same samples gives $1.1\times10^{-4}$ $[0.73,1.5]\times10^{-4}$, a factor of 8.3 below the PINN. The PINN factors of safety of the failure-region fields checked with Abaqus have an RMS error of 3.1\,\%, whereas the Bishop values are systematically higher (median ratio of Abaqus to Bishop 0.961) (Fig.~\ref{fig:estimators}d).

\begin{table}[htbp]
\centering
\caption{Probability of failure at $\cb=7$\,kPa.}
\label{tab:pf7}
\small
\resizebox{\textwidth}{!}{%
\begin{tabular}{lccc}
\toprule
Method & PINN solutions & $\Pf$ & 95\,\% interval \\
\midrule
PINN, crude MC (1 failure) & 1000 & $1.0\times10^{-3}$ & $[0.18,5.6]\times10^{-3}$ \\
PINN, lognormal fit & 1000 & $2.4\times10^{-3}$ & -- \\
PINN, importance sampling & 300 & $9.1\times10^{-4}$ & $[6.5,12.0]\times10^{-4}$ \\
Linear surrogate (MC only), FORM & 1000 & $1.6\times10^{-3}$ & -- \\
Linear surrogate (MC + IS), FORM & 1300 & $6.7\times10^{-4}$ & -- \\
Linear surrogate (MC + IS), MC $10^6$ & 1300 & $6.9\times10^{-4}$ & $[6.4,7.4]\times10^{-4}$ \\
Neural operator, MC $4\times10^6$ & 1300 & $7.0\times10^{-4}$ & $[6.8,7.3]\times10^{-4}$ \\
Heterogeneous Bishop, importance sampling & -- & $1.1\times10^{-4}$ & $[0.73,1.5]\times10^{-4}$ \\
\bottomrule
\end{tabular}}
\end{table}

\subsection{Monte Carlo with the neural operator}\label{sec:opmc}
The operator of Section~\ref{sec:operator} is evaluated on $10^6$ new fields at $\cb=5$\,kPa and on $4\times10^6$ new fields at $\cb=7$\,kPa, and Eq.~\eqref{eq:mc} is applied to the predicted $F$ with Wilson intervals. These fields are independent of the training fields. The same samples give the exceedance probability of the crest displacement and the statistics of the failure mechanism conditional on $F<1$. The HL-RF iteration of Eq.~\eqref{eq:hlrf} is also applied to the factor-of-safety head, with the gradient obtained by automatic differentiation. At $\cb=5$\,kPa the $10^6$ operator evaluations take about 4\,s and give $\Pf=0.104$, with an interval narrower than $\pm0.001$. The estimate stabilises after about $10^4$ samples (Fig.~\ref{fig:op5}e). FORM on the operator gives $\beta=1.258$ and the same value of 0.104. Both values lie inside the PINN Monte Carlo interval but 8\,\% below the point estimate of 0.113. At $\cb=7$\,kPa the $4\times10^6$ evaluations take about 9\,s, and 2816 fields fail, which gives $\Pf=7.0\times10^{-4}$ $[6.8,7.3]\times10^{-4}$. This value lies inside the importance-sampling interval and 23\,\% below the importance-sampling estimate. At both cohesions the operator is therefore slightly on the low side. The factor-of-safety head is in effect a linear regression of $\ln F$, which tends to compress the tails of the distribution.

\begin{figure}[htbp]
\centering
\includegraphics[width=\textwidth]{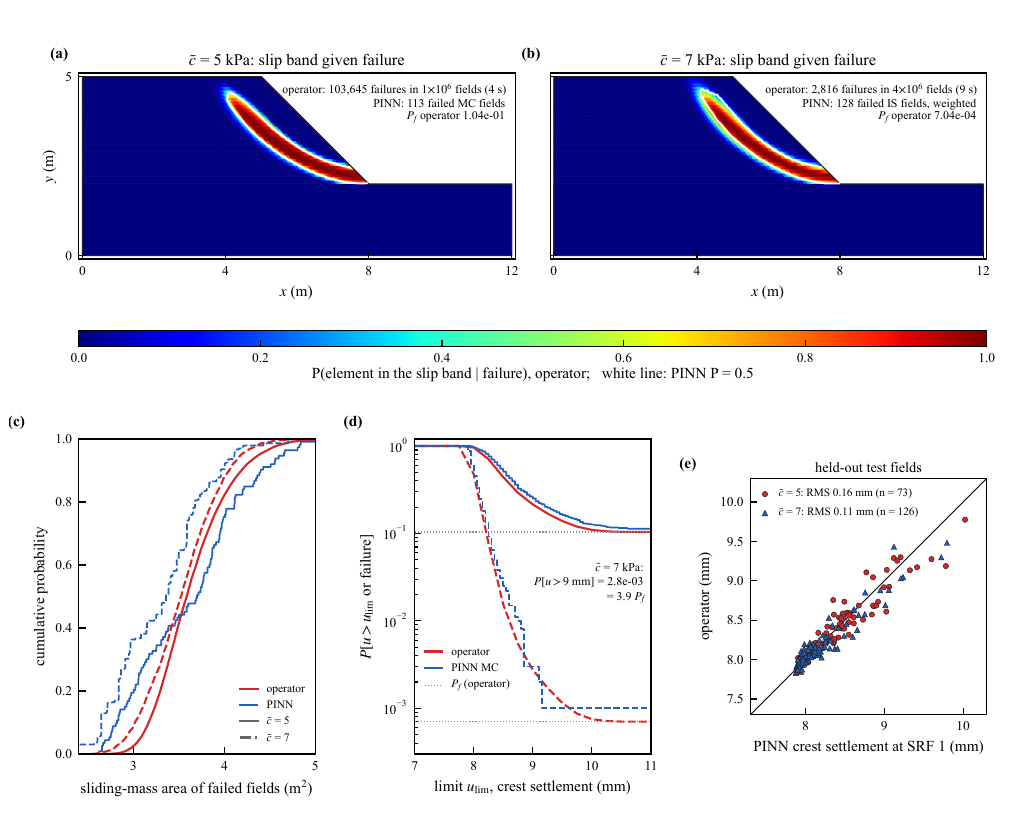}
\caption{Field-level reliability with the neural operator ($10^6$ fields at $\cb=5$\,kPa, $4\times10^6$ at $\cb=7$\,kPa). (a), (b) Probability that an element lies in the slip band given failure. The white line is the 0.5 contour of the PINN. (c) Area of the sliding mass of the failing fields. (d) Probability that the crest displacement accumulated from $s_0=0.6$ to $s=1$ exceeds $u_{\mathrm{lim}}$ or that the slope fails, with the PINN curves from the 1000 Monte Carlo fields. (e) Crest displacement at $s=1$ from the gravity state, operator against PINN on held-out test fields.}
\label{fig:fieldlevel}
\end{figure}

Beyond $\Pf$, the operator samples give quantities that require many failing fields (Fig.~\ref{fig:fieldlevel}). The probability that an element lies in the slip band given failure agrees with the PINN at both cohesions (Fig.~\ref{fig:fieldlevel}a,b), including $\cb=7$\,kPa, where the PINN failures come from the weighted importance samples. The distribution of the area of the sliding mass of the failing fields is narrower for the operator than for the PINN (Fig.~\ref{fig:fieldlevel}c), and the operator underestimates the variability of the size of the mechanism. The probability that the crest displacement accumulated between $s_0=0.6$ and $s=1$ exceeds a limit $u_{\mathrm{lim}}$, or that the slope fails, is shown in Fig.~\ref{fig:fieldlevel}d, and the accuracy of the predicted displacement on held-out fields in Fig.~\ref{fig:fieldlevel}e. At $\cb=7$\,kPa a displacement of 9\,mm is exceeded with a probability of $2.8\times10^{-3}$, 3.9 times the probability of failure. This displacement is not a physical settlement. Gravity is applied at the strengthened state $s_0=0.6$ and the stiffness is deterministic, so the quantity measures how far a field moves along an artificial weakening path, not how much the slope settles under its own weight, and its variability comes entirely from the strength field. It is reported to show that the operator returns a displacement response for every sample whose exceedance probability can be estimated at rare-event sample sizes. A serviceability analysis would require gravity applied at $s=1$ with a random stiffness, which is left to future work.

\subsection{Comparison of the estimators}\label{sec:comparison}
Fig.~\ref{fig:estimators} collects all estimates. At $\cb=5$\,kPa each estimator based on the PINN, from 500 or 1000 PINN solutions, falls within or close to the Monte Carlo interval (Fig.~\ref{fig:estimators}a). At $\cb=7$\,kPa the importance sampling, the linear surrogate trained with failure-region data and the operator agree within 25\,\% with 300 to 1300 PINN solutions (Fig.~\ref{fig:estimators}b), whereas crude Monte Carlo with the same budget cannot resolve the probability. The Bishop analysis underestimates $\Pf$ by a factor of 2.6 at $\cb=5$\,kPa and 8.3 at $\cb=7$\,kPa (Fig.~\ref{fig:estimators}c). Combining the bounds of the two 95\,\% intervals gives factors between 1.7 and 4.2 at $\cb=5$\,kPa and between 4.3 and 16 at $\cb=7$\,kPa; these ranges are conservative because both estimates use the same fields. \citet{li2016a} also found limit equilibrium estimates of $\Pf$ that differ from the finite-element estimate by a factor of about three, lower or higher depending on the slip surfaces considered in the limit equilibrium analysis. The Bishop factor of safety is higher than the PINN value by about 3\,\% at the median, similar to the difference found for the homogeneous slopes and at the design point. The factor grows as $\Pf$ decreases because an upward bias of a few percent in $F$ shifts the whole distribution, and the relative change of the exceedance probability increases towards the tail.

\section{Discussion}\label{sec:discussion}

\subsection{Computational cost}\label{sec:cost}
The PINN and the operator ran on a workstation with a 12th-generation Intel Core processor (20 virtual CPUs), 64\,GB of RAM and an NVIDIA GeForce RTX 5090 GPU (32\,GB). Abaqus ran on an Intel Core i5-7500 with four cores per job. Table~\ref{tab:cost} lists the wall-clock times.

\begin{table}[htbp]
\centering
\caption{Wall-clock times. The crude Monte Carlo times are extrapolated from the times per field.}
\label{tab:cost}
\small
\resizebox{\textwidth}{!}{%
\begin{tabular}{lll}
\toprule
Task & Size & Time \\
\midrule
Abaqus SRM & one field & 73\,s$^{a}$ \\
PINN--SRM, batch of 100 networks & 100 fields & 25--33\,min (15--20\,s per field) \\
PINN data set, $\cb=35$ / 5 / 7\,kPa & 500 / 1000 / 1300 fields$^{b}$ & 2.2 / 4.2 / 6.9\,h \\
Operator training & one data set & 0.5--1.3\,min \\
Operator: $F$, crest settlement, slip maps & $10^6$ fields & 4.2\,s \\
\multicolumn{3}{l}{Crude Monte Carlo for $\Pf=7\times10^{-4}$, COV 10\,\% ($1.4\times10^6$ fields)} \\
\quad Abaqus / PINN--SRM / operator & & 3.25 years / 325 days / 10\,s \\
\bottomrule
\end{tabular}}
\par\medskip
\begin{minipage}{\textwidth}\footnotesize
$^{a}$ Median solver time of 56 random fields (range 38--129\,s).\\
$^{b}$ Monte Carlo fields, plus 300 importance-sampling fields at $\cb=7$\,kPa. The times include a few additional fields (design points and the mean field).
\end{minipage}
\end{table}

The PINN--SRM takes 15 to 20\,s per field and Abaqus 73\,s, but the two ran on different hardware, a GPU workstation against a four-core desktop, so the per-field times are not comparable and the PINN is not proposed as a faster replacement for the finite-element analysis of one field. The benefit is that it needs no labelled data and solves many fields at once on one GPU. The main cost of the reliability analysis is the PINN data set (2.2 to 6.9\,h). For the 1300 fields of the two weak cohesions the PINN data sets took about 11\,GPU-hours, whereas sequential Abaqus runs of the same fields would take about 26\,h on the reference machine, so at this scale the total cost is comparable and the saving appears only in the operator stage. After about one minute of training, the operator evaluates $10^6$ fields in seconds. Importance sampling needs only 300 PINN solutions but gives $\Pf$ alone. The operator also gives the failure mechanism and the displacement response (Section~\ref{sec:opmc}). This ordering of cost, a few hours of PINN solutions before the operator saves anything, is the pattern that \citet{kumar2026} identifies for operator learning in geotechnics; the saving appears only when the number of required evaluations, here $10^6$, is far larger than the training set.

\subsection{Limit equilibrium and small failure probabilities}
The Bishop analysis underestimates $\Pf$ although its critical circle follows the PINN mechanism. Over 300 fields at $\cb=5$\,kPa, all critical circles pass through the toe, as does the PINN slip band (Fig.~\ref{fig:slipband}), and the two factors of safety are strongly correlated ($r=0.97$). The difference is therefore in the value of $F$, not in the shape of the slip surface. Bishop is about 3\,\% higher at the median, the usual gap between limit equilibrium and strength reduction \citep{griffiths1999}. The comparison does not separate the two ingredients of that gap, the simplified Bishop assumptions on interslice forces and the non-associated flow rule of the continuum model, because neither a rigorous non-circular limit-equilibrium method nor a run with $\psi=\varphi$ was performed. \citet{cho2010} reported ratios of 0.96 to 0.99 between strength-reduction and Bishop factors of safety for four random fields of an undrained slope.

The gap is largest in the weak fields that control $\Pf$. At $\cb=5$\,kPa the PINN-to-Bishop ratio is 0.982 for $F>1.3$ but 0.961 for $F<1.1$ (0.976 and 0.949 at $\cb=7$\,kPa). The gap also grows as the cohesion decreases: it is negligible at $\cb=35$\,kPa, and \citet{zhu2021} found close agreement for undrained slopes. The non-associated flow rule ($\psi=0$), which limit equilibrium does not represent, is a possible cause, not tested here. A finer mesh lowers the finite-element $F$ (\ref{app:mesh}) and would widen the gap.

\citet{cho2010} considered differences of this size to be of no practical consequence. Near $\Pf=10^{-3}$, however, a small error in $F$ has a large effect. Scaling the Bishop values at $\cb=7$\,kPa by 0.968 gives $\Pf=6.1\times10^{-4}$ and by 0.95 gives $1.4\times10^{-3}$, against $9.1\times10^{-4}$ from the PINN. A constant correction is therefore unreliable in the tail, consistent with the factor of about three between limit equilibrium and finite-element estimates reported by \citet{li2016a}. Limit equilibrium remains a useful guide to the failure region, as in the importance-sampling centre of Section~\ref{sec:is}, but small failure probabilities should be computed from continuum solutions.

\subsection{Neural operator versus scalar surrogates}
For $\Pf$ alone, the operator offers little beyond a simple surrogate. The factor-of-safety head is effectively a ridge regression of $\ln F$, and the ridge, Gaussian-process, and neural-network surrogates of Section~\ref{sec:form} give similar reliability indices at $\cb=5$\,kPa (1.23, 1.24, and 1.13). \citet{he2020} likewise found a linear support-vector regression sufficient. The added value is the full response of every sample. From the same $10^6$ fields, the operator gives the slip-band probability conditional on failure, the size of the sliding mass, and the exceedance probability of the crest displacement along the path (Section~\ref{sec:opmc}), which a surrogate of $F$ cannot show, although that displacement is not a physical settlement. Like any surrogate, the operator is accurate in the tail only when its training data cover the failure region. Without the importance samples, the linear surrogate gives $1.6\times10^{-3}$ instead of $6.7\times10^{-4}$ (Table~\ref{tab:pf7}). The same caution applies to deep surrogates trained on Monte Carlo samples only \citep{li2026}. The operator was not trained on Abaqus solutions of the same fields, so whether PINN labels give a better or worse operator than finite-element labels remains untested; the benefit claimed here is independence from a solver and from labelled data, not a demonstrated gain in accuracy or cost.

\subsection{Limitations}
The factor of safety of the PINN--SRM depends on the solver budget (\ref{app:budget}). The budget used here was fixed on the homogeneous slopes and verified on 52 random fields. A larger budget converges the plastic strain better but detects failure earlier, and $F$ falls 4 to 6\,\% below Abaqus at $\cb=5$ and 7\,kPa. The pipeline therefore contains no finite-element solution, but its answer is set by a finite-element calibration, and the calibration error is amplified in the tail. By the lognormal fit of Section~\ref{sec:mc}, a uniform reduction of $F$ by 4 to 6\,\% would raise $\Pf$ at $\cb=5$\,kPa from 0.11 to about 0.24 to 0.32, and by the linear surrogate of Section~\ref{sec:is}, taking the mean of $\ln F$ at $\cb=7$\,kPa as 2\,\% below the homogeneous value as at $\cb=5$\,kPa, it would raise $\Pf$ at $\cb=7$\,kPa by a factor of about four to nine. The failure probabilities reported here are therefore those of the verified budget, and a step control based on convergence rather than a fixed number of iterations is needed before they can be read as budget-independent.

The agreement in $F$ may partly reflect compensating errors. The plastic strain in the band of the weak slopes is about one quarter of the Abaqus value at $0.95F$, yet $F$ agrees within 1\,\%, and when the plastic flow is converged better $F$ falls below Abaqus (\ref{app:budget}). Under-converged plastic flow stiffens the band and delays the failure trigger, which may offset whatever drives the converged solver low. This was not isolated by a separate test. For the same reason the accepted states are approximate rather than exact equilibrium states: the nodal residual plateaus at about 0.17 of the mean nodal gravity load, far above the tolerance of a converged implicit finite-element step, and the factor of safety is the strength-reduction factor at which the optimiser can no longer carry the state within the budget. That this factor coincides with the Abaqus value within 2 to 3\,\% is an empirical finding for this budget and this problem, not a property guaranteed by the formulation.

The random error of the solver biases $\Pf$ upward. On the weak-soil fields the PINN scatters by 2.9\,\% RMS about Abaqus with no systematic sign, and three initialisations of the homogeneous weak slopes differ by 1.5\,\%, which is part of that scatter. Zero-mean scatter in $\ln F$ widens the distribution of $F$ and so raises the probability in the lower tail. If the whole 2.9\,\% is attributed to the PINN, deconvolving it from the lognormal fit at $\cb=5$\,kPa lowers $\Pf$ from 0.127 to 0.115, and deconvolving it from the linear surrogate at $\cb=7$\,kPa lowers $\Pf$ by about a factor of two. The PINN estimates are therefore more likely high than low in the tail, by an amount that is small at $\Pf\approx10^{-1}$ and comparable to the estimate itself at $\Pf\approx10^{-3}$.

The absolute failure probabilities refer to the mesh with $h=0.1$\,m. A finer mesh lowers $F$ of the weak soils by 2 to 4\,\% (\ref{app:mesh}), which would raise $\Pf$ by a factor of two to three at $\cb=7$\,kPa. This bias is common to the PINN, the operator and Abaqus, so it does not affect the comparisons between them, but it does widen the gap to the mesh-independent Bishop analysis, and the random-field Abaqus runs did not converge on the finest mesh, so the mesh-converged reference is not known.

The operator is trained for one mean cohesion and must be retrained for other statistics of the field. Including the statistics as inputs, as in the metamodel of \citet{liu2024}, would remove this restriction. Its factor-of-safety head is in effect a ridge regression of $\ln F$, and the operator estimates of $\Pf$ lie 8 and 23\,\% below the reference values, within their intervals, while the operator underestimates the spread of the sliding-mass size. The operator intervals reflect sampling error only. A systematic error of 2\,\% in $F$, the size of the held-out RMS error, would move $\Pf$ by about $-30$ to $+40$\,\% at $\cb=5$\,kPa and by a factor of about two in either direction at $\cb=7$\,kPa, so the sampling intervals understate the uncertainty of the operator by an order of magnitude in the tail. The operator also predicts displacements but not stresses, and the crest displacement it returns is accumulated along an artificial weakening path from $s_0=0.6$, so it is a displacement response and not a serviceability settlement (Section~\ref{sec:opmc}).

This study considers one plane-strain slope geometry, one correlation structure of the strength, deterministic stiffness, no pore pressure and a non-associated flow rule with $\psi=0$. The factors of 2.6 and 8.3 between the PINN and Bishop estimates belong to this slope and these mean cohesions, and the share of the non-associated flow rule in that difference was not tested. Other geometries, anisotropic correlation, random stiffness, a rigorous non-circular limit-equilibrium benchmark and three-dimensional slopes remain to be examined.

\section{Conclusions}\label{sec:conclusions}
A label-free PINN--SRM and a neural operator trained on its solutions were developed for the reliability analysis of $c$--$\varphi$ slopes with spatially variable strength. The introduction identified three gaps, a label-free solver that reproduces the finite-element SRM for non-associated Mohr--Coulomb slopes with spatially variable strength, a reliability pipeline for such slopes that is free of finite-element solutions, and an estimate of the failure mechanism and the displacement response conditional on rare failure, together with the unresolved cause of the difference between limit-equilibrium and finite-element failure probabilities. The conclusions are grouped accordingly.
\begin{enumerate}
\item \textbf{The label-free solver.} The PINN--SRM computes the factor of safety without labelled data by carrying the plastic history through the strength-reduction path with a non-associated flow rule. With a solver budget fixed on the homogeneous slopes, it agrees with Abaqus within 1\,\% on those slopes and within 2 to 3\,\% RMS on 52 random fields, and places the failure mechanism in the same weak zones. The agreement depends on the solver budget by 4 to 6\,\%, which by the fitted distributions would move $\Pf$ at the two weak cohesions by a factor of two to nine, and the plastic strain within the band is under-resolved, so the budget must be checked against a reference before the solver is applied to other problems.
\item \textbf{The finite-element-free pipeline: spatial variability.} Spatial variability governs the reliability of the weak slope. At $\cb=5$\,kPa the homogeneous analysis gives $F=1.136$, whereas Monte Carlo simulation of the PINN random fields gives $\Pf=0.113$.
\item \textbf{The finite-element-free pipeline: rare events.} Importance sampling centred on a Bishop-based design point estimates $\Pf=9.1\times10^{-4}$ at $\cb=7$\,kPa from 300 PINN solutions, with a coefficient of variation of 15\,\%. Surrogates of $F$ are accurate in the tail only when trained with data from the failure region, since the linear surrogate fitted to the Monte Carlo fields alone gives $1.6\times10^{-3}$.
\item \textbf{The finite-element-free pipeline: the operator.} The neural operator, trained on the PINN solutions alone, evaluates $10^6$ fields in seconds and gives $\Pf$ within the reference intervals at both cohesions, although 8 to 23\,\% below the point estimates, and its sampling intervals exclude a model error that is of the order of the estimate itself at $\Pf\approx10^{-3}$. No finite-element solution enters the pipeline at any stage. Abaqus confirmed the factors of safety of the 52 verification fields and the limit-state position of the design points, and served only as that check.
\item \textbf{The failure mechanism and displacement response given failure.} From the same $10^6$ operator samples, the probability that an element lies in the slip band given failure agrees with the PINN at both cohesions, the operator underestimates the spread of the sliding-mass size, and at $\cb=7$\,kPa a crest displacement of 9\,mm along the strength-reduction path is 3.9 times as likely as failure. These quantities are the value of the operator over a scalar surrogate, because for $\Pf$ alone it reduces to a linear model of $\ln F$. The displacement is accumulated from a strengthened gravity state with deterministic stiffness and is not a serviceability settlement; a serviceability analysis requires gravity at $s=1$ with random stiffness.
\item \textbf{The limit-equilibrium question.} The heterogeneous Bishop analysis underestimates $\Pf$ by factors of 2.6 and 8.3, with interval-bound ranges of 1.7 to 4.2 and 4.3 to 16. Its critical circles follow the PINN mechanism, so the error comes from a bias of about 3\,\% in $F$ between the simplified Bishop formulation and the non-associated strength-reduction model, not from the circular slip surface. The bias grows in the weak fields, is amplified in the tail, and is not removed by a constant correction. The share of the non-associated flow rule in the bias was not tested.
\end{enumerate}
Measured against the finite-element workflow it is meant to complement, the framework offers four benefits and carries one cost. It needs no labelled data, no commercial solver and no licence per core, because the PINN is trained on the equations alone and the constitutive model is differentiable code. It solves hundreds of realisations as one batched program on a single GPU, so the cost of a Monte Carlo set scales with the number of batches rather than with the number of fields. Its solutions feed an operator in the same framework, so a few hours of PINN runs are paid once and buy $10^6$ evaluations in seconds, where a direct finite-element Monte Carlo at $\Pf\approx10^{-3}$ would take years. Its operator is differentiable with respect to the random-field coefficients, which gives the FORM gradient directly and opens the way to inverse analysis. The cost is accuracy that depends on a solver budget fixed against a reference, a residual that does not vanish, a plastic strain that is under-resolved in the band and a random scatter of 3\,\% in $F$ that inflates a failure probability of $10^{-3}$ by up to a factor of two, so the finite-element method remains necessary as the independent check, the tail probabilities are bounded rather than exact, and the PINN is not a faster substitute for the finite-element method on a single slope.

Future work will replace the fixed solver budget with convergence-based step control and recompute the failure probabilities under it, apply gravity at $s=1$ with a random stiffness so that a serviceability limit state can be assessed, train the operator on finite-element solutions of the same fields as the control that this study lacks, add a physics loss to the operator (PINO), include the field statistics as operator inputs, and extend the method to three-dimensional slopes.

\section*{Statements and Declarations}
\paragraph{Funding} This research was supported by the `Petchra Pra Jom Klao Master's Degree' scholarship from King Mongkut's University of Technology Thonburi.
\paragraph{Competing interests} The authors have no competing interests to declare that are relevant to the content of this article.
\paragraph{Data availability} The code and the random-field realisations will be available at \url{https://github.com/AP1S1T/PiNNs-for-Slope-Stability-analysis} after acceptance.
\paragraph{Author contributions} Apisit Robjanghvad: methodology, software, implementation of the PINN--SRM and the neural operator, computations, verification runs, data analysis, visualisation, writing of the first draft. Sompote Youwai: conception, supervision, methodology, writing, review and editing. Both authors read and approved the final manuscript.

\appendix
\renewcommand{\theequation}{\Alph{section}.\arabic{equation}}
\renewcommand{\thefigure}{\Alph{section}.\arabic{figure}}
\renewcommand{\theHfigure}{\Alph{section}.\arabic{figure}}
\renewcommand{\theHequation}{\Alph{section}.\arabic{equation}}

\section{Gradient of the virtual-work loss}\label{app:gradient}
\setcounter{equation}{0}\setcounter{figure}{0}
This appendix shows that the gradient of the loss in Eq.~\eqref{eq:loss} is the finite-element nodal residual mapped to the network parameters (Eq.~\eqref{eq:lossgrad}).

\paragraph{Discrete kinematics} Let $\mathbf{U}\in\mathbb{R}^{2N_{\mathrm{nodes}}}$ collect the nodal displacements returned by the network through Eq.~\eqref{eq:ansatz}. At Gauss point $j$ of element $e$, the strain is stored as $\bm\varepsilon_j=(\varepsilon_{xx},\varepsilon_{yy},\varepsilon_{zz},\varepsilon_{xy})$ with $\varepsilon_{zz}=0$ (plane strain) and the tensorial shear component. The bilinear shape functions give $\varepsilon_{xx}=\sum_I\partial_xN_I\,u_I$, $\varepsilon_{yy}=\sum_I\partial_yN_I\,v_I$ and $\varepsilon_{xy}=\tfrac12\sum_I(\partial_yN_I\,u_I+\partial_xN_I\,v_I)$. An in-plane variant of the $\bar{B}$ method \citep{hughes1980} replaces the in-plane volumetric strain $\vartheta_j=\varepsilon_{xx}+\varepsilon_{yy}$ by the element mean
\begin{equation}
\bar\vartheta_e = \frac{\sum_{j\in e}w_j\,\vartheta_j}{\sum_{j\in e}w_j},\qquad
\bar\varepsilon_{xx} = \varepsilon_{xx}+\tfrac12\big(\bar\vartheta_e-\vartheta_j\big),\qquad
\bar\varepsilon_{yy} = \varepsilon_{yy}+\tfrac12\big(\bar\vartheta_e-\vartheta_j\big),
\label{eq:bbar}
\end{equation}
where $w_j$ is the Jacobian determinant at the Gauss point (the $2\times2$ Gauss weights are unity). Every operation in Eq.~\eqref{eq:bbar} is linear. The strain and the vertical displacement at the Gauss point are therefore
\begin{equation}
\bm\varepsilon_j = \bar{\mathbf{B}}_j\mathbf{U},\qquad v_j = \mathbf{N}_j\mathbf{U},
\label{eq:BU}
\end{equation}
with constant matrices $\bar{\mathbf{B}}_j$ and $\mathbf{N}_j$.

\paragraph{Gradient with respect to the nodal displacements} With the tensorial shear component, the double contraction is $\bm\sigma:\bm\varepsilon=\sigma_{xx}\varepsilon_{xx}+\sigma_{yy}\varepsilon_{yy}+\sigma_{zz}\varepsilon_{zz}+2\sigma_{xy}\varepsilon_{xy}$. The stress $\bm\sigma_j$ in Eq.~\eqref{eq:loss} comes from the return map of \ref{app:constitutive} and is held fixed during differentiation. Substituting Eq.~\eqref{eq:BU} into Eq.~\eqref{eq:loss} gives
\begin{equation}
\frac{\partial\mathcal{L}}{\partial\mathbf{U}} = \sum_j w_j\big(\bar{\mathbf{B}}_j^{\top}\mathbf{W}\bm\sigma_j+\gamma\,\mathbf{N}_j^{\top}\big) = \mathbf{f}^{\mathrm{int}}-\mathbf{f}^{\mathrm{ext}} = \mathbf{R},
\label{eq:residual}
\end{equation}
where $\mathbf{W}=\operatorname{diag}(1,1,1,2)$ accounts for the shear term. The first sum is the internal force vector of the $\bar{B}$ element. The body force $\mathbf{b}=(0,-\gamma)$ gives the external force vector $\mathbf{f}^{\mathrm{ext}}=-\gamma\sum_jw_j\mathbf{N}_j^{\top}$, which is the second sum with opposite sign. Setting $\delta\mathbf{u}$ in Eq.~\eqref{eq:vw} to each nodal shape function yields $\mathbf{R}=\mathbf{0}$ at the free degrees of freedom. Because the stress is held fixed, $\mathcal{L}$ is linear in $\mathbf{U}$ and its value has no minimum at the solution. For an elastic state the gradient coincides with that of the potential energy. For the non-associated flow rule no such potential exists, and only the gradient carries the equilibrium condition.

\paragraph{Gradient with respect to the network parameters} The chain rule gives
\begin{equation}
\frac{\partial\mathcal{L}}{\partial\bm\theta} = \Big(\frac{\partial\mathbf{U}}{\partial\bm\theta}\Big)^{\!\top}\mathbf{R}.
\label{eq:chain}
\end{equation}
The hard boundary conditions of Eq.~\eqref{eq:ansatz} set the prescribed displacements to zero for every $\bm\theta$. The corresponding rows of $\partial\mathbf{U}/\partial\bm\theta$ vanish, and only the residual at the free degrees of freedom enters Eq.~\eqref{eq:chain}. A stationary point $\partial\mathcal{L}/\partial\bm\theta=\mathbf{0}$ implies $\mathbf{R}_{\mathrm{free}}=\mathbf{0}$ when $\partial\mathbf{U}/\partial\bm\theta$ has full row rank on the free degrees of freedom, i.e. when the network can change every free nodal displacement independently. This condition is not guaranteed for a network of finite size. The residual plateau of Fig.~\ref{fig:residual} is the practical measure of how far the trained network is from it.

\paragraph{Residual monitor} The residual in Fig.~\ref{fig:residual} is evaluated with the same expression: the gradient of $\sum_jw_j[\bm\sigma_j:\bm\varepsilon_j(\mathbf{Z})+\gamma v_j(\mathbf{Z})]$ with respect to an auxiliary nodal field $\mathbf{Z}$ gives $\mathbf{R}$ by Eq.~\eqref{eq:residual}. The reported value is
\begin{equation}
\frac{\|\mathbf{R}\|_{\mathrm{rms}}}{\bar f},\qquad
\|\mathbf{R}\|_{\mathrm{rms}} = \Big(\frac{1}{n_{\mathrm{free}}}\sum_{i\in\mathrm{free}}R_i^2\Big)^{1/2},\qquad
\bar f = \frac{\gamma A}{N_{\mathrm{nodes}}},
\label{eq:monitor}
\end{equation}
with $A=\sum_jw_j$ the area of the domain and $n_{\mathrm{free}}$ the number of free degrees of freedom.

\section{Constitutive update}\label{app:constitutive}
\setcounter{equation}{0}\setcounter{figure}{0}
\paragraph{Invariants} With $p=-\tfrac13\operatorname{tr}\bm\sigma$ (compression positive) and the deviatoric stress $\mathbf{s}=\bm\sigma+p\mathbf{I}$,
\begin{equation}
J_2 = \tfrac12\,\mathbf{s}:\mathbf{s},\qquad q=\sqrt{3J_2},\qquad J_3=\det\mathbf{s},\qquad \cos3\Theta=\frac{27J_3}{2q^3},
\label{eq:invariants}
\end{equation}
with $\Theta\in[0,\pi/3]$. The plane-strain components are $J_3=s_{xx}s_{yy}s_{zz}-s_{zz}\sigma_{xy}^2$ and $J_2=\tfrac12(s_{xx}^2+s_{yy}^2+s_{zz}^2)+\sigma_{xy}^2$. In the implementation $J_2$ is bounded below by $10^{-12}$ and $\cos3\Theta$ is bounded to $\pm(1-10^{-6})$, which avoids the singularities at the hydrostatic axis and at the corners of the Mohr--Coulomb surface.

\begin{figure}[H]
\centering
\includegraphics[width=\textwidth]{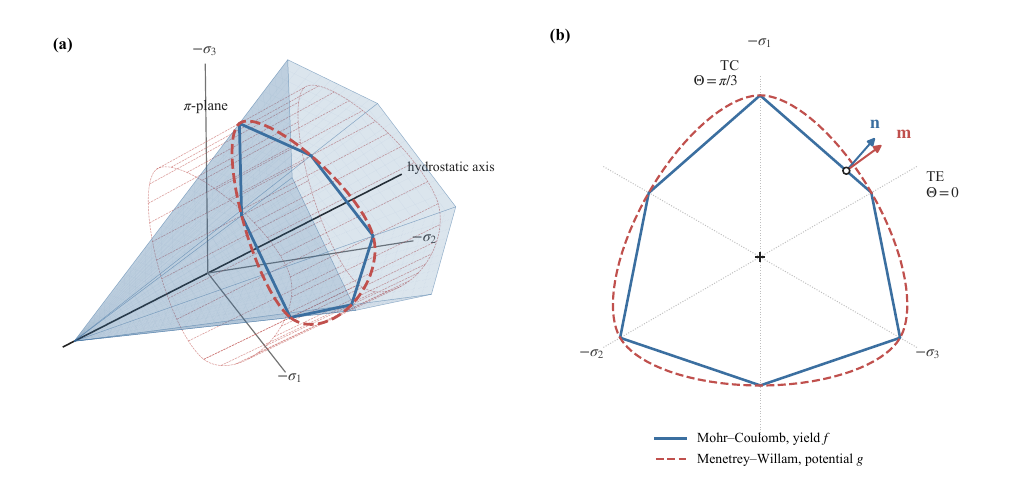}
\caption{Yield function and plastic potential for $\varphi=20^\circ$. (a) Principal stress space (compression positive): the Mohr--Coulomb pyramid and the Menetrey--Willam potential with $\psi=0$, which is parallel to the hydrostatic axis. The bold curves are their sections by one deviatoric ($\pi$) plane. (b) The same $\pi$-plane. The radius is proportional to $1/R(\Theta)$ and scaled to 1 at triaxial compression (TC). TE: triaxial extension. The arrows are the deviatoric parts of $\mathbf{n}=\partial f/\partial\bm\sigma$ and $\mathbf{m}=\partial g/\partial\bm\sigma$ at one stress point on the yield surface.}
\label{fig:surfaces}
\end{figure}

\paragraph{Yield function and plastic potential} The yield function is Eq.~\eqref{eq:yield}. The deviatoric shape of the Menetrey--Willam potential in Eq.~\eqref{eq:potential} is
\begin{equation}
\begin{aligned}
R_{mw}(\Theta,e) &= \frac{4(1-e^2)\cos^2\Theta+(2e-1)^2}{2(1-e^2)\cos\Theta+(2e-1)\sqrt{4(1-e^2)\cos^2\Theta+5e^2-4e}}\;\frac{3-\sin\varphi}{6\cos\varphi},\\
e &= \frac{3-\sin\varphi}{3+\sin\varphi},
\end{aligned}
\label{eq:rmw}
\end{equation}
with $e$ bounded to $[0.51,1]$ and $\varphi$ the reduced friction angle $\varphi_s$ of Eq.~\eqref{eq:srf}. For $\psi=0$ the potential reduces to $g=R_{mw}\,q$. Both surfaces pass through the six corners of the Mohr--Coulomb hexagon, since $R_{mw}=R_{mc}$ at $\Theta=0$ and $\pi/3$, and the potential is smooth between them (Fig.~\ref{fig:surfaces}). The flow rule is therefore non-associated in the deviatoric plane as well as in the meridional plane: away from the corners the plastic flow $\mathbf{m}$ is not normal to the yield surface.

\paragraph{Flow directions} The normals $\mathbf{n}=\partial f/\partial\bm\sigma$ and $\mathbf{m}=\partial g/\partial\bm\sigma$ follow from the chain rule through the invariants,
\begin{equation}
\begin{aligned}
\frac{\partial(\cdot)}{\partial\bm\sigma} &= \frac{\partial(\cdot)}{\partial p}\Big(-\frac{\mathbf{I}}{3}\Big)+\frac{\partial(\cdot)}{\partial q}\,\frac{3\mathbf{s}}{2q}+\frac{\partial(\cdot)}{\partial\Theta}\,\frac{\partial\Theta}{\partial\bm\sigma},\\
\frac{\partial\Theta}{\partial\bm\sigma} &= -\frac{9}{2\sin3\Theta}\Big(\frac{\operatorname{dev}(\mathbf{s}\mathbf{s})}{q^3}-\frac{3J_3}{q^4}\,\frac{3\mathbf{s}}{2q}\Big),
\end{aligned}
\label{eq:normals}
\end{equation}
where $\operatorname{dev}(\mathbf{s}\mathbf{s})=\mathbf{s}\mathbf{s}-\tfrac23J_2\mathbf{I}$ is the derivative of $J_3$. For the yield function $\partial f/\partial p=-\tan\varphi$, $\partial f/\partial q=R_{mc}$ and $\partial f/\partial\Theta=q\,\partial R_{mc}/\partial\Theta$. For the potential with $\psi=0$, $\partial g/\partial p=0$, $\partial g/\partial q=R_{mw}$ and $\partial g/\partial\Theta=q\,\partial R_{mw}/\partial\Theta$. The $\Theta$ term is set to zero where the bound on $\cos3\Theta$ is active. The closed-form expressions were compared with automatic differentiation of $f$ and $g$ at random stress states and random strengths: the median relative error is $1.5\times10^{-7}$ for both $\mathbf{n}$ and $\mathbf{m}$.

\paragraph{Cutting-plane return map} Given the total strain $\bm\varepsilon$, the plastic strain $\bm\varepsilon^p_n$ of the last accepted SRF step and the reduced strength $(c_s,\tan\varphi_s)$, the stress is computed with a fixed number of four cutting-plane iterations \citep{simo1985}:
\begin{equation}
\begin{aligned}
&\bm\sigma^{(0)} = \mathbb{D}:(\bm\varepsilon-\bm\varepsilon^p_n),\qquad \Delta\bm\varepsilon^p=\mathbf{0},\\
&\text{for }k=0,\dots,3:\quad \Delta\lambda^{(k)} = \frac{\langle f(\bm\sigma^{(k)})\rangle}{\max\big(\mathbf{n}^{(k)}:\mathbb{D}:\mathbf{m}^{(k)},\,10^{-6}\big)},\\
&\qquad \Delta\bm\varepsilon^p\leftarrow\Delta\bm\varepsilon^p+\Delta\lambda^{(k)}\mathbf{m}^{(k)},\qquad
\bm\sigma^{(k+1)} = \mathbb{D}:(\bm\varepsilon-\bm\varepsilon^p_n-\Delta\bm\varepsilon^p),
\end{aligned}
\label{eq:cuttingplane}
\end{equation}
where $\langle x\rangle=\max(x,0)$ and $\mathbf{n}^{(k)}$, $\mathbf{m}^{(k)}$ are evaluated at $\bm\sigma^{(k)}$. The stress used in the loss is $\bm\sigma^{(4)}$. The plane-strain elasticity matrix acting on $(\varepsilon_{xx},\varepsilon_{yy},\varepsilon_{zz},\varepsilon_{xy})$ is
\begin{equation}
\begin{aligned}
\mathbb{D} &= \begin{bmatrix}
\lambda_{\mathrm{L}}+2\mu & \lambda_{\mathrm{L}} & \lambda_{\mathrm{L}} & 0\\
\lambda_{\mathrm{L}} & \lambda_{\mathrm{L}}+2\mu & \lambda_{\mathrm{L}} & 0\\
\lambda_{\mathrm{L}} & \lambda_{\mathrm{L}} & \lambda_{\mathrm{L}}+2\mu & 0\\
0 & 0 & 0 & 2\mu
\end{bmatrix},\\
\lambda_{\mathrm{L}} &= \frac{E\nu}{(1+\nu)(1-2\nu)},\qquad \mu=\frac{E}{2(1+\nu)},
\end{aligned}
\label{eq:Dmatrix}
\end{equation}
and the double contraction $\mathbf{n}:\mathbb{D}:\mathbf{m}$ uses the weight $\mathbf{W}$ of \ref{app:gradient}. The update is evaluated without gradient tracking, consistent with the stop-gradient stress of Eq.~\eqref{eq:loss}. When a trial step is accepted, $\bm\varepsilon^p_n$ is replaced by $\bm\varepsilon^p_n+\Delta\bm\varepsilon^p$.

\section{Discrete Karhunen--Lo\`eve expansion}\label{app:kl}
\setcounter{equation}{0}\setcounter{figure}{0}
The eigenpairs of Eq.~\eqref{eq:kl} solve the Fredholm equation $\int_\Omega\rho(\mathbf{x},\mathbf{x}')\,\phi(\mathbf{x}')\,\dd\mathbf{x}'=\lambda\,\phi(\mathbf{x})$ with the kernel of Eq.~\eqref{eq:markov}. The Nystr\"om method \citep{betz2014} replaces the integral by a quadrature with the element centroids $\mathbf{x}_e$ as integration points and the element areas $A_e$ as weights, which gives $\mathbf{C}\mathbf{A}\bm\phi=\lambda\bm\phi$, where $C_{ef}=\rho(\mathbf{x}_e,\mathbf{x}_f)$ and $\mathbf{A}=\operatorname{diag}(A_e)$. The substitution $\mathbf{v}=\mathbf{A}^{1/2}\bm\phi$ gives the symmetric problem
\begin{equation}
\mathbf{A}^{1/2}\mathbf{C}\mathbf{A}^{1/2}\mathbf{v}_k = \lambda_k\mathbf{v}_k,\qquad
\bm\phi_k = \mathbf{A}^{-1/2}\mathbf{v}_k,\qquad
\sum_e A_e\,\phi_k(\mathbf{x}_e)\,\phi_l(\mathbf{x}_e) = \delta_{kl},
\label{eq:nystrom}
\end{equation}
from which the $M=100$ largest eigenpairs are kept. The sign of each mode is fixed by making the entry of largest magnitude positive, which makes $\bm\xi$ reproducible across machines. The truncated expansion underestimates the point variance \citep{betz2014}, $\hat\sigma^2(\mathbf{x}_e)=\sum_k\lambda_k\phi_k^2(\mathbf{x}_e)<1$, and each field is therefore rescaled to unit variance,
\begin{equation}
g(\mathbf{x}_e) = \frac{1}{\hat\sigma(\mathbf{x}_e)}\sum_{k=1}^{M}\sqrt{\lambda_k}\,\xi_k\,\phi_k(\mathbf{x}_e).
\label{eq:rescale}
\end{equation}
Two independent vectors $\bm\xi_c,\bm\xi'\in\mathbb{R}^{100}$ give $g_c$ and $g'$, and the friction-angle field follows as $g_\varphi=\rho_{c\varphi}\,g_c+\sqrt{1-\rho_{c\varphi}^2}\,g'$. The two fields share the same eigenpairs, and only the KL variables are cross-correlated \citep{cho2010}. The KL vector of a realisation is $\bm\xi=(\bm\xi_c,\bm\xi')\in\mathbb{R}^{200}$, and the element values follow from Eq.~\eqref{eq:lognormal}. The same field can be evaluated at other points, for example the centroids of a finer mesh, by the Nystr\"om interpolation formula
\begin{equation}
\phi_k(\mathbf{x}) = \frac{1}{\lambda_k}\sum_e\rho(\mathbf{x},\mathbf{x}_e)\,A_e\,\phi_k(\mathbf{x}_e),
\label{eq:nystrominterp}
\end{equation}
which returns $\phi_k$ exactly at the original centroids.

\section{Mesh convergence of the reference solutions}\label{app:mesh}
\setcounter{equation}{0}\setcounter{figure}{0}
The factor of safety was recomputed with Abaqus on meshes with $h=0.2$, 0.1 and 0.05\,m, with the random fields evaluated at the new centroids by Eq.~\eqref{eq:nystrominterp} (Fig.~\ref{fig:mesh}). The coarse mesh is within 2\,\% of the paper mesh ($h=0.1$\,m) in all cases. For the homogeneous slopes, refinement to $h=0.05$\,m changes $F$ by $+0.1$, $-2.0$ and $-3.7$\,\% at $\cb=35$, 5 and 7\,kPa. For the random fields, the implicit solver stops on the finest mesh before a mechanism forms, and the last converged strength reduction factor is not a factor of safety. With automatic stabilisation on both meshes, the finest mesh is 2.6\,\% below the paper mesh for field 90, consistent with the homogeneous results. Stabilisation itself raises $F$ by 11\,\% and is not used for reference values.

The mesh bias of about $-2$ to $-4$\,\% for the weak soils is common to the PINN, the operator and Abaqus, and does not affect the comparisons of Sections~\ref{sec:verif}--\ref{sec:reliability}. The absolute failure probabilities refer to $h=0.1$\,m. A uniform reduction of $F$ by 2 to 4\,\% would raise $\Pf$ at $\cb=5$\,kPa from 0.11 to 0.19--0.25 and $\Pf$ at $\cb=7$\,kPa by a factor of two to three.

\begin{figure}[H]
\centering
\includegraphics[width=\textwidth]{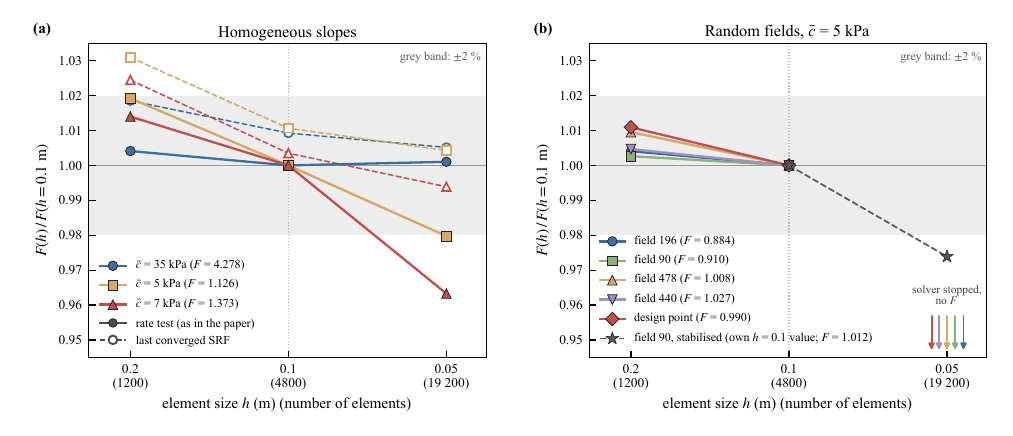}
\caption{Mesh convergence of the Abaqus factor of safety. (a) Homogeneous slopes, $F(h)$ normalised by the value on the paper mesh ($h=0.1$\,m). Filled markers: factor of safety from the displacement-rate criterion; open markers: last converged strength reduction factor. (b) Five random fields at $\cb=5$\,kPa. On $h=0.05$\,m the implicit solver stopped before a mechanism formed (arrows), and no factor of safety is available. Stars: field 90 with automatic stabilisation, normalised by its own value on $h=0.1$\,m. The grey band is $\pm2$\,\%.}
\label{fig:mesh}
\end{figure}

\section{Sensitivity to the solver budget}\label{app:budget}
\setcounter{equation}{0}\setcounter{figure}{0}
The homogeneous slopes were solved again with plastic steps of 1.4 or 0.7\,\% of the SRF, 1000 or 3000 iterations per step and two stricter failure rules (Fig.~\ref{fig:budget}). With the budget of the paper (1.4\,\%, 1000 iterations), $F$ lies within about 2\,\% of Abaqus at all three cohesions. With 0.7\,\% and 3000 iterations the fields converge better: at $\cb=5$\,kPa the error of the plastic strain falls from 83 to 44\,\% and the correlation with Abaqus rises from 0.51 to 0.90. The failing trials, however, then appear below the Abaqus limit, and $F$ is 4 to 6\,\% lower than Abaqus at $\cb=5$ and 7\,kPa. At $\cb=35$\,kPa $F$ does not depend on the budget. The factor of safety of the weak slopes therefore depends on the budget by a few percent, and the agreement with Abaqus holds for the budget used in the paper.

\begin{figure}[H]
\centering
\includegraphics[width=\textwidth]{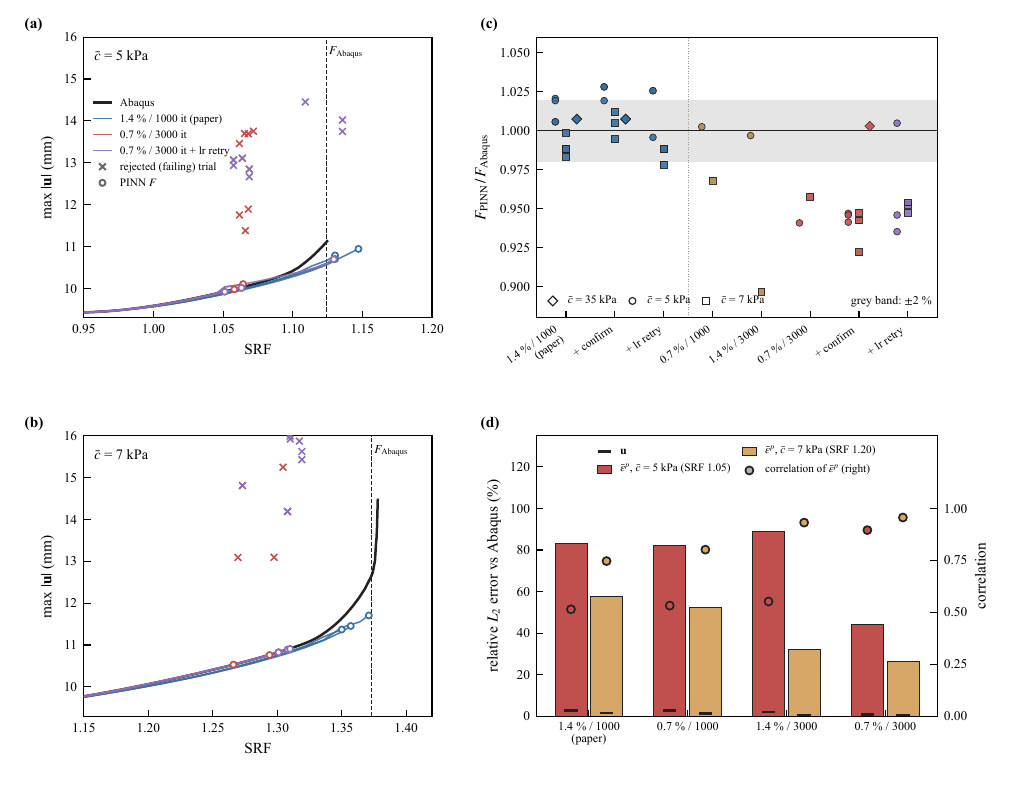}
\caption{Sensitivity of the PINN--SRM to its solver budget (homogeneous slopes). (a), (b) Maximum displacement along the SRF path near the limit: Abaqus and the accepted PINN states of three budgets. Crosses are rejected (failing) trials. (c) $F_{\mathrm{PINN}}/F_{\mathrm{Abaqus}}$ for every budget and failure rule. (d) Field errors at a common SRF below the limit.}
\label{fig:budget}
\end{figure}

\end{document}